\documentclass[aps,pra,10pt,twocolumn,a4paper,nofootinbib,preprintnumbers,superscriptaddress]{revtex4-2}
\usepackage{CJK}
\usepackage{amsmath,amsfonts,amssymb,amsthm}
\usepackage{eqnarray}
\usepackage{mathcomp}
\usepackage{srcltx}
\usepackage[normalem]{ulem} 
\usepackage{graphics,graphicx}
\usepackage{dcolumn}
\usepackage{bm}
\usepackage{hyperref}
\usepackage{url}
\usepackage[mathlines]{lineno}
\usepackage[dvipsnames]{xcolor}
\usepackage{upgreek}
\usepackage{color, colortbl}
\usepackage{array}
\usepackage{cleveref} 
\usepackage[acronym]{glossaries}
\usepackage{xfrac}
\usepackage{siunitx}
\DeclareSIUnit\pulse{pulse}
\usepackage[version=4,arrows=pgf-filled,textfontname=sffamily,mathfontname=mathsf]{mhchem}
\normalsize 
\usepackage{multirow, tabularx, ragged2e, makecell} 
\usepackage{booktabs} 
\usepackage{fancyhdr}

\newcolumntype{L}[1]{>{\raggedright\let\newline\\\arraybackslash\hspace{0pt}}m{#1}}
\newcolumntype{C}[1]{>{\centering\let\newline\\\arraybackslash\hspace{0pt}}m{#1}}
\newcolumntype{R}[1]{>{\raggedleft\let\newline\\\arraybackslash\hspace{0pt}}m{#1}}

\def\({\left(}
\def\){\right)}
\def\[{\left[}
\def\]{\right]}

\newcommand{\beq}{\begin{equation}}
\newcommand{\eeq}{\end{equation}}
\newcommand{\bea}{\begin{eqnarray}}
\newcommand{\eea}{\end{eqnarray}}

\definecolor{LightGray}{gray}{0.9}
\usepackage{subfigure} 

\newcommand{\supp}{Supplemental Material}

\newacronym{apd}{APD}{avalanche photo diode}
\newacronym{awg}{AWG}{arbitrary waveform generator}
\newacronym{bs}{BS}{beam splitter}
\newacronym{cir}{CIR}{circulator}
\newacronym{cw}{CW}{continuous waveform}
\newacronym{cwdm}{CWDM}{coarse wavelength division multiplexing}
\newacronym{dc}{DC}{direct current}
\newacronym{dfb}{DFB}{distributed feedback}
\newacronym{dlcz}{DLCZ}{Duan-Lukin-Cirac-Zoller}
\newacronym{dwdm}{DWDM}{dense wavelength division multiplexing}
\newacronym{edfa}{EDFA}{erbium doped fiber amplifier}
\newacronym{epc}{EPC}{electronic polarization controller}
\newacronym{fim}{FIM}{fast intensity modulation}
\newacronym{fpga}{FPGA}{field-programmable gate array}
\newacronym{Imod}{Int. Mod.}{intensity modulator}
\newacronym{ld}{LD}{laser diode}
\newacronym{lidt}{LIDT}{laser-induced damage threshold}
\newacronym{mcf}{MCF}{multi-core fiber}
\newacronym{mdi}{MDI}{measurement-device-independent}
\newacronym{oil}{OIL}{optical injection locking}
\newacronym{osa}{OSA}{optical spectrum analyzer}
\newacronym{otdr}{OTDR}{optical time-domain reflectometry}
\newacronym{pbs}{PBS}{polarizing beam splitter}
\newacronym{pc}{PC}{polarization controller}
\newacronym{pd}{PD}{photodiode}
\newacronym{ph mod}{Ph. Mod.}{phase modulator}
\newacronym{pmpqkd}{PMP-QKD}{post-measurement pairing QKD}
\newacronym{qber}{QBER}{quantum bit error rate}
\newacronym{qc}{QC}{quantum communications}
\newacronym{qkd}{QKD}{quantum key distribution}
\newacronym{rf}{RF}{radio frequency}
\newacronym{snspd}{SNSPD}{superconducting nanowire single photon detectors}
\newacronym{skc0}{SKC\textsubscript{0}}{repeaterless secret key capacity bound}
\newacronym{skc1}{SKC\textsubscript{1}}{single repeater secret key capacity bound}
\newacronym{skr}{SKR}{secret key rate}
\newacronym{smf}{SMF}{single mode fiber}
\newacronym{tf}{TF}{twin-field}
\newacronym{tfqkd}{TF-QKD}{twin field QKD}
\newacronym{tha}{THA}{Trojan-horse attack}
\newacronym{twirl}{TWIRL}{Trojan-wavelength in the reference light}
\newacronym{voa}{VOA}{variable optical attenuator}
\newacronym{wcp}{WCP}{weak coherent pulse}

\usepackage{fancyhdr}
\fancypagestyle{plain}{%
  \fancyhf{}
  \fancyhead[C]{\small Phys.\ Rev.\ A \textbf{113}, 032613 (2026) --- \href{https://doi.org/10.1103/71m5-3c5n}{https://doi.org/10.1103/71m5-3c5n}}
  \fancyfoot[C]{\thepage}
}

\begin{document}

\graphicspath{ {./images/} }

\title{Reference-Beam Attacks against Twin-Field Quantum Key Distribution using Optical Injection Locking}

\author{Sergio Ju\'arez}
\thanks{These two authors contributed equally to this work.}
\affiliation{Toshiba Europe Limited, 208 Cambridge Science Park, Cambridge CB4 0GZ, UK}
\affiliation{Escuela de Ingeniería de Telecomunicación, Department of Signal Theory and Communications, University of Vigo, Vigo E-36310, Spain}
\author{Alessandro Marcomini}
\thanks{These two authors contributed equally to this work.}
\affiliation{Vigo Quantum Communication Center, University of Vigo, Vigo E-36310, Spain}
\affiliation{Escuela de Ingeniería de Telecomunicación, Department of Signal Theory and Communications, University of Vigo, Vigo E-36310, Spain}
\affiliation{AtlanTTic Research Center, University of Vigo, E-36310, Spain}
\author{Mikhail Petrov}
\affiliation{Vigo Quantum Communication Center, University of Vigo, Vigo E-36310, Spain}
\affiliation{Escuela de Ingeniería de Telecomunicación, Department of Signal Theory and Communications, University of Vigo, Vigo E-36310, Spain}
\affiliation{AtlanTTic Research Center, University of Vigo, E-36310, Spain}
\author{Robert~I.~Woodward}
\affiliation{Toshiba Europe Limited, 208 Cambridge Science Park, Cambridge CB4 0GZ, UK}
\author{Toby~J.~Dowling}
\affiliation{Toshiba Europe Limited, 208 Cambridge Science Park, Cambridge CB4 0GZ, UK}
\author{R. Mark Stevenson}
\affiliation{Toshiba Europe Limited, 208 Cambridge Science Park, Cambridge CB4 0GZ, UK}
\author{Marcos Curty}
\affiliation{Vigo Quantum Communication Center, University of Vigo, Vigo E-36310, Spain}
\affiliation{Escuela de Ingeniería de Telecomunicación, Department of Signal Theory and Communications, University of Vigo, Vigo E-36310, Spain}
\affiliation{AtlanTTic Research Center, University of Vigo, E-36310, Spain}
\author{Davide Rusca}
\affiliation{Vigo Quantum Communication Center, University of Vigo, Vigo E-36310, Spain}
\affiliation{Escuela de Ingeniería de Telecomunicación, Department of Signal Theory and Communications, University of Vigo, Vigo E-36310, Spain}
\affiliation{AtlanTTic Research Center, University of Vigo, E-36310, Spain}

\begin{abstract}
   Twin-Field Quantum Key Distribution (TF-QKD) has become a leading protocol to bring quantum communications to the national scale. The protocol requires the establishment of a shared phase and frequency reference between distant parties, which is commonly achieved by using an external reference laser in an Optical Injection Locking (OIL) architecture. In this work, we analyze the side channels in OIL-based TF-QKD that may arise from adversarial manipulation of the various degrees of freedom of this untrusted reference beam. We experimentally demonstrate two realistic attack scenarios: fast intensity modulation of the reference laser, and additional signals embedded in the reference light exploiting wavelengths undetectable by conventional monitoring techniques. These attacks can allow a potential eavesdropper to deterministically increase the mean photon number of the sources, or circumvent the decoy-state technique, respectively. To counter these vulnerabilities, we propose practical and highly effective countermeasures that reinforce the security of TF-QKD systems without significant additional complexity or performance degradation.
\end{abstract}

\maketitle
\thispagestyle{fancy}

\section*{Introduction}

\Gls*{qkd} enables two legitimate parties, Alice and Bob, to share a symmetric encryption key that can be used to achieve information-theoretically secure communications. In particular, it exploits the laws of quantum mechanics to ensure that any eavesdropping attempt by a malicious external party (Eve) cannot go undetected~\cite{BB84,Ekert.1991,Pirandola.2020, lo2014secure}.

To date, \gls*{qkd} has matured into a well-established technology whose application has been demonstrated on the level of inter-city networks~\cite{sasaki2011field, martin2024madqci, chen2025implementation}. Next-generation protocols such as \gls*{tfqkd}, which utilize single-photon interference, offer significantly improved resilience against channel losses~\cite{Lucamarini.2018,Wang.2018,Xu.2020, curty2019simple}, effectively doubling the achievable communication distances~\cite{curty.2009,Pirandola.2017}, reaching up to 1000~km point-to-point~\cite{Minder.2019,Pittaluga.2021,Chen.2022,Wang.2022,Liu.2023}. These properties make \gls*{tfqkd} particularly attractive for the realization of nationwide quantum networks~\cite{pittaluga2025long,Chen.2021,Liu.2021,Clivati.2022,Zhou.2023}.

Nevertheless, the broader adoption of \gls*{qkd} still faces some challenges, particularly the need to guarantee the security of practical implementations, since real devices inevitably deviate from the idealized assumptions. These deviations must be accounted for, either through improved security proofs or through enhancements in physical hardware, a domain collectively known as implementation security~\cite{Xu.2020b, zapatero2025implementation}. Unaddressed device imperfections can lead to unintended side channels, resulting in undetected information leakage which may compromise the overall system security~\cite{jain2016attacks, Makarov.2024, bsi2023implementation}.

In this regard, \gls*{tfqkd} possesses a crucial advantage, as it belongs to the \gls*{mdi} class of protocols. \Gls*{mdi} protocols remove all security assumptions from the receiver, ensuring security even if Eve completely controls the measurement apparatus~\cite{Lo.2012}. Consequently, implementation security efforts can be entirely focused on the transmitter side.

Practical implementations of \gls*{tfqkd} require a shared phase and frequency reference between Alice's and Bob's encoders. Typically, this reference is provided by the untrusted central measurement node (Charlie), which distributes a classical reference beam through a dedicated service channel. The received beam stabilizes the laser sources via optical phase-locked loops (OPLLs) or, more commonly, \gls*{oil}~\cite{Ye.2003,Comandar.2016c, paraiso2021advanced}. \Gls*{oil} is particularly favoured due to its inherent stability, practicality, and lack of necessity for active feedback control~\cite{pittaluga2025long,du2024twin}. However, permitting an external beam, potentially altered by Eve en route, to directly enter the encoders raises critical concerns regarding implementation security~\cite{peng2025practical,Lucamarini.2015,Makarov.2024}.

In this work, we provide the first thorough analysis of implementation security specifically tailored to the \gls*{oil} architecture used in \gls*{tfqkd}. We investigate potential attack scenarios enabled by manipulation of the reference laser, examining multiple optical degrees of freedom and revealing two new potential side channels. Additionally, we also provide practical countermeasures for them.

The first identified side channel exploits a \gls*{fim} of the reference laser, performed on the same timescale as the pulse encoding or faster, which temporarily increases the emitted photon number without detection by conventional monitoring. 

The second side channel exploits back-reflections from the \gls*{ld} in the \gls*{oil} configuration, and the limited spectral response of monitoring devices. The separate channel that goes into the encoders allows Eve to embed an additional signal at a \gls*{twirl}. When this undetected light reaches the encoder, it could possibly leak the complete information about the intensity and phase settings used in the quantum signals. Note that this channel cannot be blocked without defeating its purpose.

Our analysis underscores the necessity for careful spectral filtering of the injected reference beam and highlights the importance of employing monitoring systems (``watchdogs") that track the timing and integrity of the incoming signals with high temporal resolution. Implementing these countermeasures effectively closes these side channels, significantly strengthening the security of \gls*{oil} \gls*{tfqkd} implementations.

The remainder of this paper is structured as follows: first, we provide a detailed examination of potential attack vectors originating from Eve’s manipulation of Charlie’s reference signal, identifying FIM and out-of-band \gls*{twirl} signals as realistic threats. Next, we experimentally demonstrate the feasibility of these attacks and evaluate suitable countermeasures. Finally, we present our conclusions and broader implications of our findings for the practical security of \gls*{tfqkd}.

\section*{Potential side channels in OIL TF-QKD}

\begin{figure}[t]
\centering
\includegraphics[width=\linewidth]{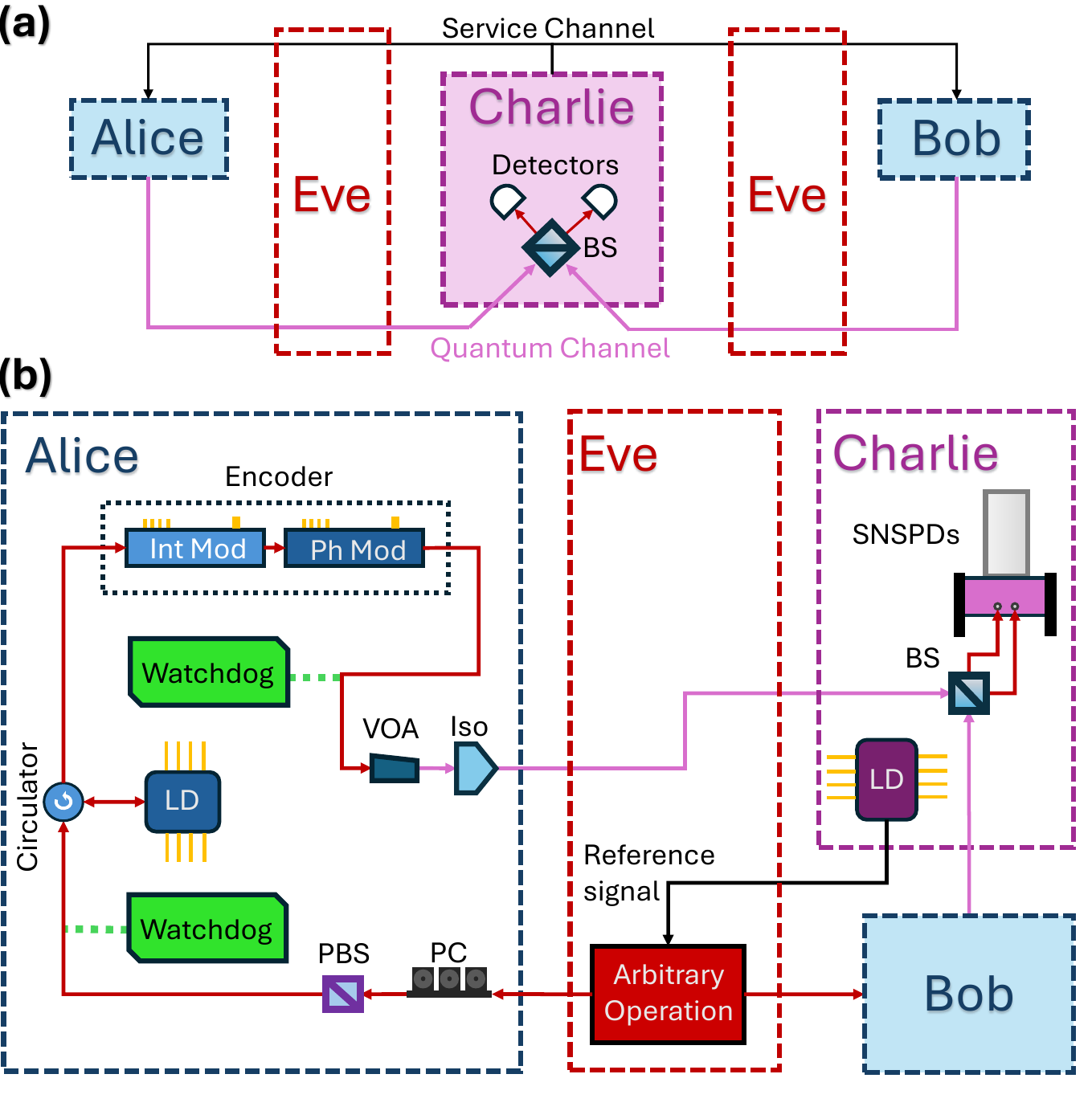}
\caption{
    \textbf{Schematics of a basic \gls*{tfqkd} system and encoder architecture based on \gls*{oil}. (a)} Overview of a typical \gls*{tfqkd} setup where Charlie distributes a coherent reference via the service channel. Alice and Bob encode quantum signals and send them back through the quantum channel for interference at Charlie’s \gls*{bs} and measurement. Eve can manipulate both channels and, in the most extreme case, take full control of Charlie. \textbf{(b)} Key components of an \gls*{oil}-based encoder. The reference passes through a \gls*{pc} and \gls*{pbs} for single-polarization injection, then locks the local \gls*{ld} via a circulator. The output is encoded using an intensity modulator (Int Mod) and phase modulator (Ph Mod), attenuated to the single-photon level with a \gls*{voa}, and sent to Charlie through an isolator (Iso). Interference and detection are performed at Charlie using a \gls*{bs} and single-photon detectors such as \gls*{snspd}s. The watchdogs guarantee the integrity of these signals. Bob’s setup is identical to Alice’s.}
\label{fig:BasicTF_setup}
\end{figure}

In Fig.~\ref{fig:BasicTF_setup}(a), we present a simplified schematic of a typical \gls*{tfqkd} system consisting of Alice, Bob, and Charlie. The central node (Charlie) provides a coherent optical reference to the transmitters (Alice and Bob) through the service channel. Alice and Bob individually prepare quantum states, which are subsequently sent back to Charlie via the quantum channel for single-photon interference and measurement. Both channels can potentially be manipulated by Eve, and, in the worst case, Charlie himself could be entirely under Eve’s control. Fig.~\ref{fig:BasicTF_setup}(b) details the most important components of an encoder and receiver within an \gls*{oil}-based \gls*{tfqkd} architecture. The optical reference received from Charlie first passes through a \gls*{pc} and a \gls*{pbs}, ensuring single-polarization injection to optimize the injection-locking efficiency. The optical reference goes then through a circulator and is injected into the encoder's laser, locking its emission with that of Charlie's laser. This results in a shared optical frequency and a fixed phase offset, which enables coherent operation across all sources. Subsequently, the locked signal passes through the encoder, which uses intensity and phase modulators to craft the train of pulses through pulse carving and manual (active) phase randomization. Finally, the pulses get attenuated to the single-photon level using a \gls*{voa} and are transmitted back to Charlie after passing through an isolator that prevents attacks from the quantum channel against the transmitter.

\begin{figure*}[t]
\centering
\includegraphics[width=\textwidth]{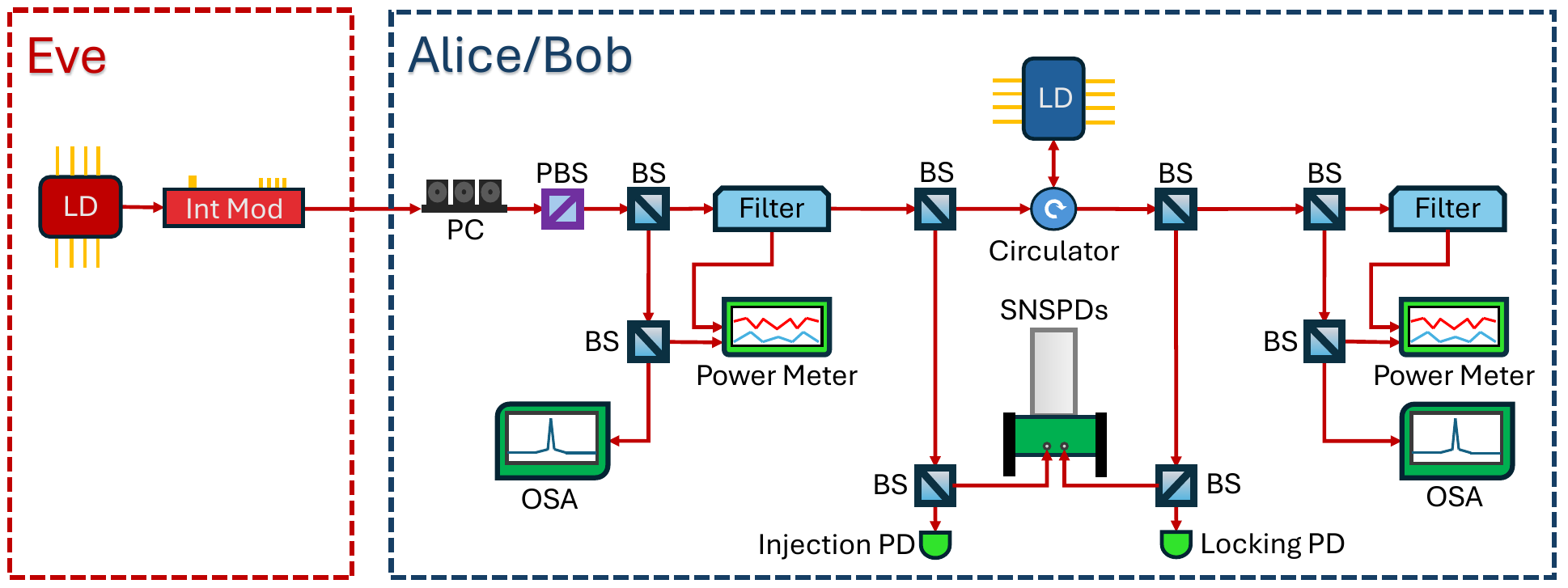}
\caption{
    \textbf{Experimental setup for studying the FIM attack.}    Here, Eve is assumed to fully control the intensity of the reference laser, implemented using an intensity modulator (Int Mod). The modulated reference signal enters Alice (or Bob), passing through a PC and a PBS, then undergoes spectral filtering to isolate the agreed reference from extrinsic wavelengths before injection-locking the LD. The behavior of the injected and locked signals is characterized by SNSPDs and an Optical Spectrum Analyzer. The signals are also monitored by two PDs with GHz detection bandwidths, one to observe the signal injected into the encoder’s laser (Injection PD) and another to monitor the laser’s output after locking (Locking PD), and a power meter that checks filtered and unfiltered signals before and after the locking. After the second filter, the signal would normally pass through the encoder, be attenuated, and then sent to Charlie, as in Fig.~\ref{fig:BasicTF_setup}(b).
    }
\label{fig:setups}
\end{figure*}

\begin{figure*}[t]
\centering
\includegraphics[width=\textwidth]{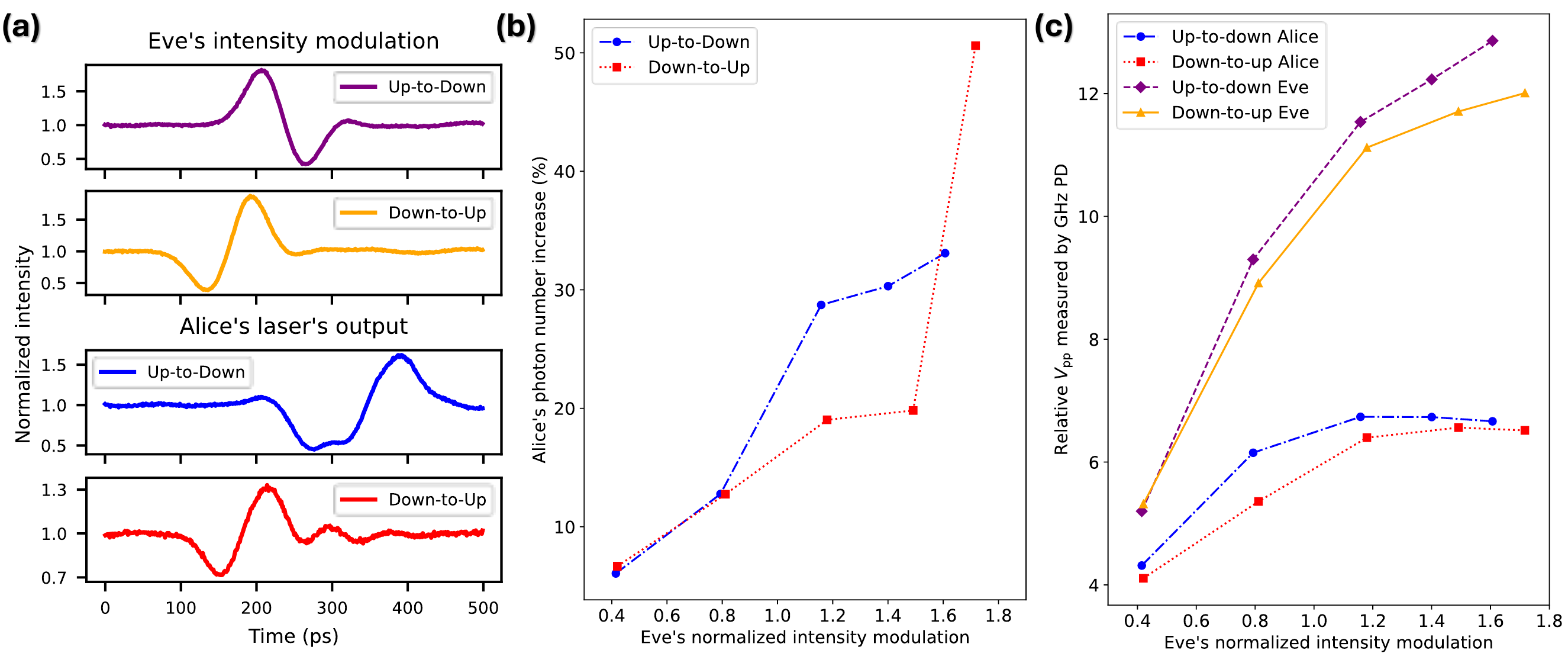}
    \caption{
     \textbf{Detection performance of watchdog detectors under \gls*{fim} attacks.} We study the attack under various amplitude modulations: ``Eve's normalized intensity modulation" in all sub-figures (and in Fig.~\ref{fig:TrojanWL_results}) refers to the difference in power between the minima and maxima of the modulated laser signal (peak-to-peak), relative to the baseline power of the unmodulated signal. \textbf{(a)} Examples of the two primary intensity modulation patterns tested: ``Up-to-Down" (power increased, then reduced) and ``Down-to-Up" (power reduced, then increased), both maintaining constant average intensity. \textbf{(b)} \gls*{snspd} measurements showing a transient increase in mean photon number (up to $51$\%) during the modulation intervals. Slow-integrating power meters failed to detect these modulations. The abrupt increase at the final point of the “Down-to-Up” configuration is due to the merging between the main pulse and the trailing ringing. \textbf{(c)} fast-\gls*{pd} output reliably identifying all modulation types tested, enabling immediate attack detection and limiting adversarial interference to a denial-of-service scenario.
    }
    \label{fig:Watchdog_results}
\end{figure*}

This architecture simplifies frequency and phase stabilization, ensuring coherent interference at Charlie’s measurement node. However, adversaries might attempt to exploit any of the various degrees of freedom of the injected reference signal, such as polarization, phase, intensity, or wavelength (frequency), to gain information through side-channel attacks.

Polarization-based attacks are mitigated by the \gls*{pbs} at the input of the security perimeter, which transmits a single polarization component while suppressing the orthogonal one with a typical extinction ratio of 30~dB. As a result, polarization attacks relying on rapid fluctuations of the polarization of the reference beam lose one projection, while the remaining accepted component only fluctuates in intensity. Consequently, polarization attacks become equivalent to intensity attacks. Phase manipulation attacks are ineffective due to the active phase randomization in Alice's and Bob's encoders after the \gls*{oil} stage, and any excessive phase fluctuations are detectable as they will increase the \gls*{qber}, resulting at most in a denial-of-service scenario. 

Attacks targeting the reference laser's intensity could  allow Eve to induce transient increases in the output power of Alice’s and Bob’s lasers, implying a threat to the single-photon level encoding of \gls*{qkd}. For instance, increasing the intensity of the coherent state $\langle\mu\rangle=0.4\to0.6$ doubles the multi-photon emission probability. A \gls*{fim} attack that goes undetected would thus enable Eve to circumvent the security provided by the no-cloning theorem and the decoy state method~\cite{Lo.2005, sajeed.2015, huang.2019, Wang.2018}. 

Spectral attacks encompass two distinct scenarios: manipulating the frequency of the injected reference itself, and injecting additional optical signals at wavelengths far removed from the intended reference wavelength. While the former approach inevitably worsens the interference at the central node, thus enlarging the \gls*{qber}, the latter exploits vulnerabilities arising from wavelength-dependent responses of the detectors and variations in attenuation across optical components within the system. Importantly for \gls*{tfqkd} systems, using a service channel to introduce a reference into the system creates an entry point for \gls*{twirl} signals that cannot be fully closed.

To mitigate these potential attacks, we recommend the use of monitoring detectors (``watchdogs") strategically placed within the encoder setup, as illustrated in Fig.~\ref{fig:BasicTF_setup}(b). A first watchdog placed before optical injection ensures the integrity of the incoming reference signal, while a second watchdog positioned immediately after the encoder monitors the output pulse train before attenuation to the single-photon level, enabling classical-level signal processing. These watchdogs should continuously monitor power, timing, and spectral characteristics, triggering alerts upon detecting anomalies indicative of hardware malfunctions or potential eavesdropping attempts.

However, the design of these watchdogs requires careful consideration, as Eve could exploit their intrinsic limitations to evade detection~\cite{sajeed.2015}. For instance, slow \glspl*{pd} that integrate optical power over relatively long time intervals, compared to the encoding rate, might be sufficient to treat slow intensity fluctuations, but they will fail to detect FIM. Eve could maintain a constant average optical power, lowering and increasing the power within time windows shorter than the integration period, thus avoiding detection. Similarly, \glspl*{pd} with responsivity optimized for specific wavelengths (typically aligned with the intended communication wavelength) can be vulnerable to wavelength-dependent \glspl*{tha}. Eve could impersonate Charlie by injecting 
anomalous signals at wavelengths outside the \gls*{pd}’s detection range. In such scenarios, the watchdog would register only minimal increases in optical power, despite the presence of a high-intensity \gls*{twirl} signal. This extrinsic signal could then propagate undetected through Alice’s and Bob’s setups, surviving the wavelength-dependent attenuation of critical optical components such as \gls*{voa}s, intensity modulators, and phase modulators. Consequently, the \gls*{twirl} signal could reach deep into the encoders, extracting sensitive information about the encoded quantum states (including intensity setting, basis choice, and even bit choice) and leaking it back to Eve.

To accurately determine the required performance characteristics of these watchdogs and enhance their resilience, we experimentally recreated these attack scenarios under controlled conditions. Using detectors with different temporal and spectral responses, we evaluated the effectiveness of these attacks and established robust design criteria for the watchdogs, as presented in the following section. 

\section*{Experimental demonstration and countermeasures}

\subsubsection*{Fast intensity modulation attack}

In Fig.~\ref{fig:setups} we present the setup used to investigate the FIM attack, where Eve modulates only the intensity of the reference laser. Such signal is routed into Alice’s (equivalently, Bob’s) station, passing through a \gls*{pc} and a \gls*{pbs} to ensure a single polarization, effectively converting polarization-based attacks into intensity-based ones. A narrowband spectral filter is applied to reroute unwanted spectral components (see below) and isolate the expected reference wavelength, preventing injection of extrinsic light. The spectrally filtered signal is then used to injection-lock Alice's laser, synchronizing its phase and frequency with the modulated reference. Importantly, all lasers operate in continuous wave (CW) mode, and the encoder and the rest of the components shown in Fig.~\ref{fig:BasicTF_setup}(b) are omitted in the setup shown in Fig.~\ref{fig:setups}. This is because the experiment focuses specifically on assessing how \gls*{fim} of the reference signal affects the \gls*{oil} process, potentially compromising transmitter security.

\begin{figure*}[t]
\centering
\includegraphics[width=1\linewidth]{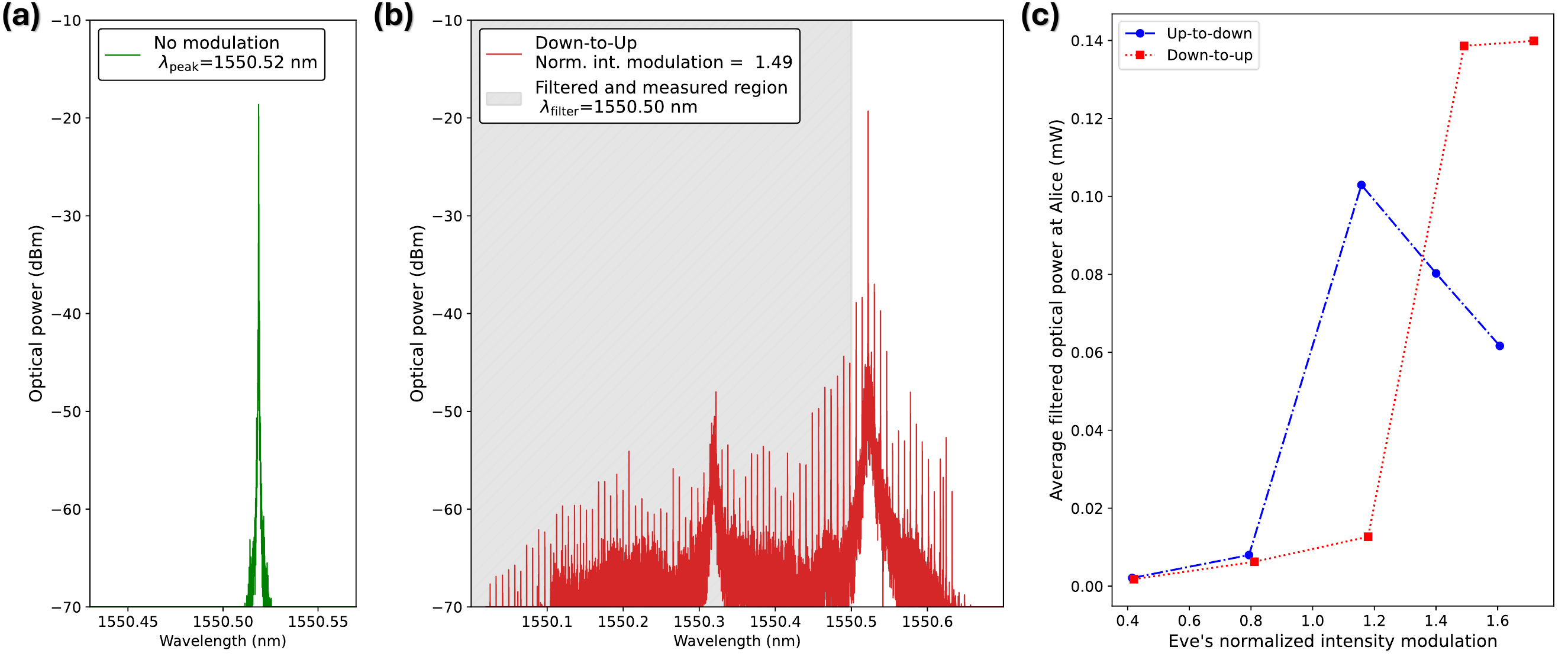}
    \caption{
    \textbf{Spectral detection of \gls*{fim} attacks. (a)} Reference optical spectrum of Alice’s \gls*{ld} without modulation. \textbf{(b)} Spectrum under active \gls*{fim} at $1$ GHz repetition rate, showing prominent sidebands around the main wavelength. \textbf{(c)} Power detected in the filtered sidebands region using a slow ($25$~$\upmu$s integration) power meter. The observed increase in power clearly indicates modulation-induced spectral broadening, providing a simple yet effective secondary detection method. The decay observed towards the end results from variations in the sidebands moving outside the tunable filter's passband; these features are also visible on the right side of the spectrum.
    }
    \label{fig:Filters_results}
\end{figure*}

Superconducting nanowire single-photon detectors (\gls*{snspd}s) equipped with a time tagger, together with an \gls*{osa}, are used to characterize the properties of the modulated reference signal prior to the locking of Alice's laser, and also, to study the effects that the attack has on the effective photon numbers and the spectra of the locked laser. Using slow, integrating power meters, we monitor the power levels of both the unfiltered optical signal and, in conjunction with the filters, any contributions from extrinsic or unexpected wavelengths, measured both before and, more importantly, after the locking stage.

The first option we investigate for a potential watchdog consists of two GHz-bandwidth fast-\glspl*{pd} monitoring the high-speed behavior of the system: the Injection PD tracks the temporal profile of the signal arriving at the encoder's laser via the circulator, while the Locking PD observes the output of the laser after the locking process. To assess both the threat posed by \gls*{fim} attacks and the effectiveness of these fast-\gls*{pd} as countermeasures, we experimentally characterized an injection-locked laser under attack conditions. Specifically, to resolve a 1~GHz modulation of the reference laser and its impact on the encoding laser, we employed fast-\glspl*{pd} with bandwidths of 10~GHz and 40~GHz, respectively.

The detection results of an injection locked laser under a \gls{fim} attack (Fig.~\ref{fig:Watchdog_results}(a)) are displayed in Figs.~\ref{fig:Watchdog_results}(b)~and~\ref{fig:Watchdog_results}(c). The \gls*{snspd}s data in Fig.~\ref{fig:Watchdog_results}(b) indicates that Eve's modulation can produce transient increases in the mean photon number by up to $51$\% during the modulated time intervals. This demonstrates that the \gls*{fim} attack is a credible threat, as it enables a controlled increase in the photon number emitted by Alice or Bob. Conventional power meters register no modulation when integrating in their fastest configuration of $25$~$\upmu$s. The results of these measurements are in the \supp \ \cite{ThisPaperSM}. Although we observe power meters to be oblivious to these modulations, we note that even if they could detect sustained increases or decreases in average power, Eve could strategically time the modulated pulses to dilute the fluctuations, making them indistinguishable from detector noise. Therefore, our analysis shows that slow-integrating power meters alone are insufficient against this type of attack.

The effectiveness of this attack, and the failure of slow-integrating power meters to detect it, can be understood from fundamental physical constraints on Eve's strategy. First, to affect the photon number of an encoded pulse, Eve's modulation must occur within the temporal window of that pulse. This confines any effective attack to timescales on the order of the pulse duration, or shorter. Furthermore, to evade detection by slow-integrating power meters, Eve must maintain a constant average optical power. This requirement forces her to employ symmetric modulation patterns such as ``Up-to-Down'' or ``Down-to-Up'' (see Fig.~\ref{fig:Watchdog_results}(a)), where any transient increase in power is compensated by a corresponding decrease. As a consequence, the total excess energy that Eve can redistribute into any given pulse is limited by the product of the nominal reference power and the pulse duration. An extensive overview of the pulse designing phase, together with additional modulation scenarios, can be found in the Methods. Eve is also constrained by the physics of \gls*{oil}. In fact, Eve does not have direct control over Alice's laser output: she can only modulate the injected reference signal, and the laser's output response is mediated by the injection locking dynamics. These dynamics (including the finite locking bandwidth, the cavity photon lifetime, and amplitude-phase coupling in the gain medium) impose a minimum response time that prevents arbitrarily fast modulations of the locked laser output, regardless of how rapidly Eve modulates her input.

Taking into account these constraints on the attack, a fast-PD with bandwidth matching (or exceeding) the system clock rate and with a rise time comparable to (or shorter than) the pulse duration can resolve intensity variations within each pulse period, and will therefore detect Eve's manipulation. Crucially, if Eve attempts to evade detection by compressing her modulation into shorter timescales, she faces a fundamental trade-off: shorter modulation windows proportionally reduce the energy she can modulate per pulse. In other words, the same strategy that might help Eve avoid detection simultaneously renders her attack ineffective.

As reported in Fig.~\ref{fig:Watchdog_results}(c), the fast-\glspl*{pd} reliably detect anomalies for every modulation tested, promptly identifying the attack, enabling countermeasures, and limiting Eve's impact to a denial-of-service scenario since once the attack is detected the protocol must be interrupted.

Furthermore, owing to amplitude–phase coupling in the laser's gain medium, oscillations in the injected power convert partly into instantaneous frequency shifts, resulting into both intensity modulations and spectral sidebands~\cite{paraiso2021advanced}. This spectral distortion in Alice's local laser output provides another signature of malicious intervention, which can be detected as a countermeasure. To this end, the combined use of the slow-integrating power meter alongside spectral filters to detect unexpected wavelength components results in a secondary detection venue against \gls*{fim} attacks.

The measurement results of the power meter aided by the spectral filters are shown in Fig.~\ref{fig:Filters_results}. Comparing an unmodulated spectral baseline (Fig.~\ref{fig:Filters_results}(a)) with an active modulation scenario (Fig.~\ref{fig:Filters_results}(b)), clear sidebands emerge. A narrowband spectral filter directs these otherwise absent spectral sidebands into a power meter, providing a straightforward detection method even with slow integration times. The results of Fig.~\ref{fig:Filters_results}(c) were taken with the $25$~$\upmu$s integration of the power meter, but are observable with integration times up to $100$~$\upmu$s. This approach enables reliable identification of \gls*{fim} attacks by monitoring unexpected spectral shifts, enabling Alice and Bob to abort the QKD session for protection, since any increase in power in these wavelengths would be created by an attack on the encoding laser.

For practical implementations, we recommend a calibration procedure in which the injected reference is modulated at various depths while monitoring the out-of-band power through a narrowband filter positioned just outside the unperturbed laser's spectral range. This establishes, for the specific laser and OIL configuration, the minimum modulation depth producing detectable spectral distortion, ensuring the watchdog threshold is appropriately set.

\begin{figure}
    \centering
    \includegraphics[width=\columnwidth]{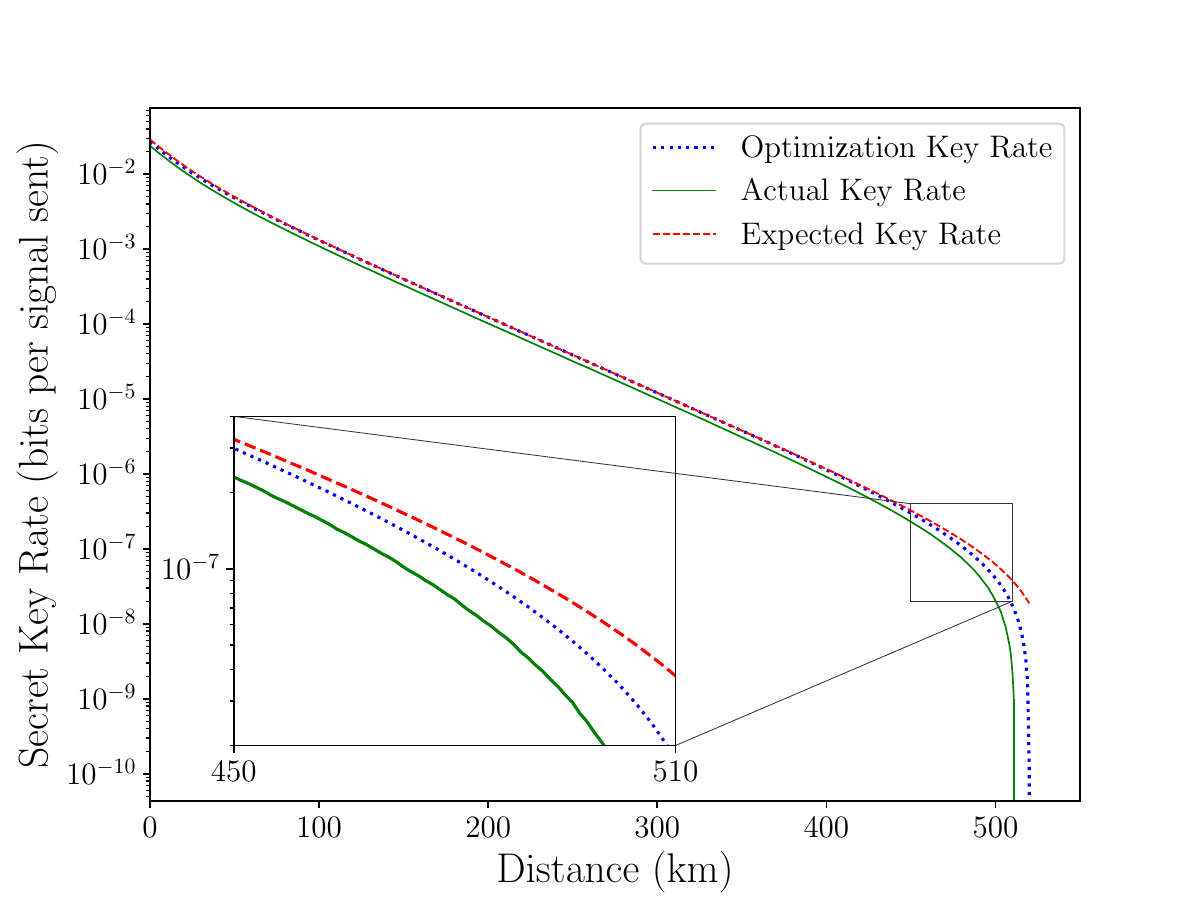}
    \caption{\textbf{\gls*{fim} attacks lead to an erroneous estimation of the \gls*{skr}.} 
    To run their protocol, Alice and Bob must first optimize some parameters, based on their system characteristics and channel conditions. This optimization yields a theoretical estimate for the expected \gls*{skr} (blue dotted curve). However, under Eve's \gls*{fim} attack the signal states they actually prepare have enlarged intensity. Here we set the enhancement factor to $+51\%$, corresponding to the maximum intensity increase experimentally measured in Fig.~\ref{fig:Watchdog_results}(b). If Alice and Bob are aware of the intensity increment, they can correctly apply the security analysis, obtaining a lower bound on the \gls*{skr} that accurately reflects the attack (green solid curve). Nevertheless, if they were oblivious to the attack and employed the expected intensity value in the \gls*{skr} calculation together with the actual statistics observed, they would retrieve an erroneous \gls*{skr} which is higher that the actual one provided by the proof (red dashed line), thus overestimating it. Further details and simulation parameters are provided in ``Methods".}
    \label{fig:SKR}
\end{figure}

Crucially, underestimating the actual optical power emitted by Alice and Bob, even if still under the single-photon regime, can lead to an overestimation of the achievable SKR. To illustrate this, we consider the ``Sending-or-Not-Sending" TF-QKD protocol, and compute the asymptotic \gls*{skr} in the presence of the attack (see Ref.~\cite{Wang.2018} for the protocol description and security proof). The results of this analysis are presented in Fig.~\ref{fig:SKR}, with details about the simulations provided in Methods. Plots in Fig.~\ref{fig:SKR} demonstrate that, even in the asymptotic limit, there is a noticeable discrepancy in the \gls*{skr} estimation under attack conditions. Such discrepancies are expected to become more pronounced in finite-key scenarios, since one has to heavily rely on measurements in the control basis to provide meaningful estimates of the phase error rate, and since this procedure is based on studying the interference of the weak coherent pulses sent by Alice and Bob, the \gls*{fim}-induced amplification of such pulses will result in out-of-ordinary estimates of the protocol's parameters, potentially severely impacting performance.

While a full protocol-specific security analysis accounting for finite-size effects is beyond the scope of this work, we note that in extreme cases with small data block sizes, the induced discrepancy between the expected and the actual \gls*{skr} could even lead to the former surpassing the theoretical upper bound on the achievable key rate, yielding a completely insecure~\cite{huang.2019,curty.2009} estimate. We cannot evaluate such claim for the case at hand, as no explicit upper bound on the \gls*{skr} for \gls*{tfqkd} is available, to be best of our knowledge. Although this means that no definitive insecurity claims can be done at present time, it has to be anticipated that a carefully designed attack could increase the emitted intensity beyond the values in Fig.~\ref{fig:Watchdog_results}(b), and the upper bound on the \gls*{skr} will thus eventually be violated. 
Therefore, accurately bounding the emitted states’ intensity is critical to ensure practical security.

\subsubsection*{Trojan-wavelength in the reference light attack}

We now move on to characterize the \gls*{twirl} attack in the context of \gls*{oil} \gls*{tfqkd}, where we experimentally investigate the spectral response of the laser cavity itself. Critically, any Trojan-wavelength light which bypasses the watchdogs and gets reflected by the laser cavity will be injected into the encoder and undergo the same encoding steps, potentially leaking full information about the intensity and phase settings if not effectively filtered. It should be noted that this component arrangement is unique to \gls*{tfqkd}, as the reference signal reaches the \gls*{ld} via the circulator connected to the service channel, rather than from the quantum channel.

\begin{figure*}[t]
\centering
\includegraphics[width=1\linewidth]{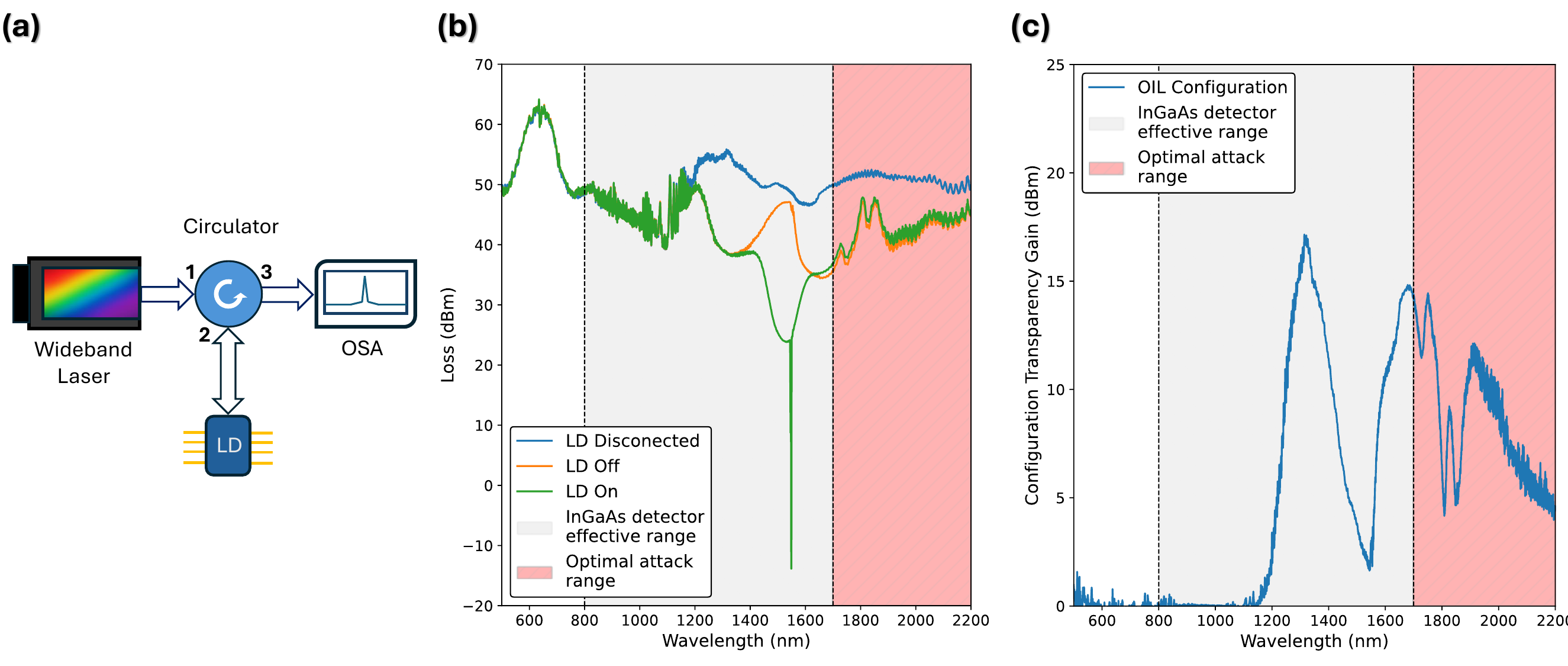}
    \caption{\textbf{Spectral characterization of laser cavity reflections relevant to \gls*{twirl} attacks.}
    \textbf{(a)} Experimental setup for studying the \gls*{twirl} attack. A wideband laser and an \gls*{osa} are employed to characterize the spectral response of the \gls*{ld} over a broad wavelength range from 500 nm to 2200 nm.
    \textbf{(b)} Measured spectral losses for different experimental configurations of the encoder setup. Losses were determined by subtracting the spectra obtained with the wideband laser alone from those measured when the circulator and \gls*{ld} were connected. Three configurations are compared: circulator alone (no \gls*{ld} connected), circulator with \gls*{ld} connected but powered off, and circulator with \gls*{ld} connected and powered on. \textbf{(c)} Effective transparency gained in the whole Circulator-\gls*{dfb} \gls*{ld} configuration, computed as the difference between the losses with the \gls*{ld} disconnected and the \gls*{ld} connected but off. Results show significant deviation of the expected transparency due to reflections when the cavity is connected, and allow to determine the optimal wavelengths for the attack. 
    }
    \label{fig:TrojanWL_results}
\end{figure*}

Previous studies have documented the spectral behavior of common encoder components such as attenuators and modulators~\cite{tan2025wide}. However, the response of a laser cavity under illumination by wavelengths far outside its typical operating band has not yet been characterized, particularly in the context of \gls*{oil} architectures. As this is the fundamental missing piece to properly compute the total attenuation along the full optical path experienced by light at every wavelength, our setup (depicted in Fig.~\ref{fig:TrojanWL_results}(a)) isolates the laser cavity using only a fiber-optic circulator and a \gls*{dfb} laser, deliberately omitting other encoder components in order to specifically analyze cavity interactions at out-of-band wavelengths. In doing so, we aim to reveal any previously unidentified spectral dependencies or unexpected vulnerabilities inherent in \gls*{oil}-based \gls*{tfqkd} architectures.

 To explore a broad spectral region, we use a supercontinuum wideband fiber laser covering a wavelength range of $500$-$2200$~nm. While all encoder components are optimized for the telecom C-band centered around $1550$~nm, we intentionally illuminate the \gls*{dfb} laser with wavelengths far from its nominal injection locking region. Importantly, the laser under test does not have an isolator and is not locking to this out-of-band light. Reflected signals, largely bypassing the cavity due to off-resonance conditions, are collected via the circulator and measured using an \gls*{osa}. This arrangement allows us to directly measure and characterize how the laser cavity structure affects signals at wavelengths significantly outside its normal operational range, identifying spectral regions where unexpectedly high levels of reflected power could aid Eve’s \gls*{twirl} attack.

It is worth noting that while our experimental characterization was performed using a \gls*{dfb} laser (the most commonly used laser type within \gls*{qkd} systems), the \gls*{twirl} attack is expected to apply broadly to any laser used in OIL architectures. This generality arises because \gls*{oil} inherently requires an optical pathway into the laser cavity, and this same pathway is accessible to light at \gls*{twirl} wavelengths. Any laser cavity will exhibit some degree of reflectivity at out-of-band wavelengths via interaction with cavity interfaces or facets, particularly since the gain medium provides neither absorption nor amplification far from the operating wavelength. The quantitative reflectivity spectrum will depend on the specific laser design, and therefore a thorough security assessment should include characterization of the laser employed in each implementation.

Figure~\ref{fig:TrojanWL_results}(b) shows the spectral losses observed by comparing spectra from the wideband source alone to those from three different experimental configurations: circulator only (no \gls*{ld} connected), circulator with \gls*{ld} connected but powered off, and circulator with \gls*{ld} actively lasing. Conversely, Fig.~\ref{fig:TrojanWL_results}(c) highlights the effective transmissivity gained through the presence of the \gls*{ld} cavity in the \gls*{oil} configuration, calculated by comparing configurations with the \gls*{ld} connected and disconnected (\gls*{ld} off only). This approach reveals the ideal wavelengths to use for this attack. That is, those with positive transmissivity gains that are undetectable by typical watchdogs (like InGaAs \glspl*{pd}, which are optimized for $\sim1550$~nm and sharply lose responsivity beyond $1700$~nm~\cite{rogalski2011infrared}), thus allowing the injected signals to bypass the cavity unnoticed.

The results in Fig.~\ref{fig:TrojanWL_results}(c) clearly indicate significant cavity-induced reflections within $1200$~nm to $2200$~nm. Eve could exploit this broad spectral range to inject a high-power optical signal, with a wavelength significantly above what an InGaAs detector would detect. Under these conditions, the detected optical power would substantially underestimate the actual injected power, allowing Eve’s \gls*{twirl} signal to propagate through the encoder at classical power levels, despite the typical attenuation mechanisms implemented within quantum encoding setups.

These results illustrate how the \gls*{twirl} attack must be carefully taken into account for the security of any \gls*{oil} based \gls*{tfqkd} system, as it allows to completely learn the encoding settings when overlooked. Fortunately, to prevent this attack one only requires an adequate amount of spectral filtering with high enough attenuation at these external wavelengths, thus ensuring that for optical frequencies outside of the locking range, the intensity of the light after the encoder is sufficiently weak to guarantee security. 

In order to compute the required amount of attenuation one must understand what is the maximum amount of power that Eve could inject into Alice and Bob's devices, as well as what is the maximum acceptable output power of the encoders, according to security proofs. For the former is typically considered, as a worse-case scenario, the so-called \gls*{lidt}. For light at telecom wavelengths, Ref.~\cite{Lucamarini.2015} reports a theoretical estimate of the \gls*{lidt} value of $55$kW, while in a recent certification guideline Ref.~\cite{Makarov.2024} considers as a more reasonable conservative assumption that the maximum possible insertion power for a realistic Eve is $100$W. We remark that this power threshold only applies to fiber, and does not take into consideration possible damage to connectors and other components, meaning that in practice it is probably still over-conservative. 

Crucially, the value of the \gls*{lidt} strongly depends on the wavelength~\cite{Carr.2003}. Therefore, by assuming a \gls*{lidt} of $100$ W at $\lambda_0 = 1550$~nm, the \gls*{lidt} at an arbitrary wavelength $\lambda$ can be found by following the square-root dependence as
\begin{equation}\label{eqn: LIDT equation}
    \text{LIDT}(\lambda) = \sqrt{\frac{\lambda}{\lambda_0}}\times 100 \text{ W}.
\end{equation}
Note that Eq.~\ref{eqn: LIDT equation} suggests that the maximum allowed input power for Eve is larger for longer wavelengths, yielding approximately 105~W at 1700~nm. Within the encoder, the VOA typically provides 70--75~dB of attenuation at the operating wavelength to reach single-photon levels; however, this attenuation is wavelength-dependent, and components optimized for 1550~nm may provide substantially less isolation at out-of-band \gls*{twirl} signals \cite{tan2025wide}. When considering  the worst case scenario, it is standard to assume that Eve can bypass channel losses entirely, placing herself right outside Alice's security perimeter. However, for an attacker operating from Charlie's position, the attack signal would face additional fiber losses at out-of-band wavelengths due to infrared absorption (approximately 0.75~dB/km at 1700~nm versus $\sim$0.2~dB/km at 1550~nm~\cite{Palais2005}). Noting the differences in loss across wavelengths, it is more conservative to provide suppression of unexpected out-of-band wavelengths inside the encoder, assuming the \gls*{lidt} threshold.

The \gls*{twirl} attack is conceptually equivalent to the standard \gls*{tha} coming from the quantum channel. Therefore, to produce a quantitative and theoretically secure upper bound for the maximum tolerable output power a comprehensive analysis against these, specific to \gls*{tfqkd}, is required. Unfortunately, to date such analysis has yet to be established, meaning that no formal estimation of the induced information leakage is currently available for this scheme. One can get a rough estimate of the required isolation from security proofs for other \gls*{qkd} schemes, under the assumption that the maximum allowed output power of a \gls*{tfqkd} transmitter in the Trojan mode is comparable to that of prepare-and-measure \gls*{qkd}. For these, an overall input-to-output optical isolation of approximately $200$~dB is typically considered necessary to approach the performance of an ideal, leakage-free scenario~\cite{Lucamarini.2015, Curras-Lorenzo.2025a, Tamaki.2016, Wang.2018.leaky, Navarrete.2022, curras-lorenzo.2025b, sixto2025quantum}. Fortunately, such levels of attenuation can be achieved using off-the-shelf spectral filters. When cascaded, multiple units can provide the required out-of-band suppression with less than 10~dB of total in-band loss, while completely suppressing \gls*{twirl} signals~\cite{tan2025wide}. 

In any case, we remark that to properly close this security loophole for a given system, one should consider the maximum tolerable Trojan light intensity established by a dedicated security proof, and design the attenuation scheme accordingly.

\section*{Conclusions}

In this work, we have performed a detailed analysis of all possible degrees of freedom of the reference laser that could pose a security threat in \gls*{oil}-based \gls*{tfqkd} systems. We identified two new viable attack vectors: \gls*{fim} of the injected reference, and Trojan signals at wavelengths well outside the optimal detection range of usual watchdogs, equipped with sufficient power to pass through the system despite existing attenuation.

To counter external \gls*{fim} of the generated pulse train, we recommend the use of a fast-\gls*{pd} as a watchdog, ideally as fast as the encoding itself, in order to take advantage of the fact that the reference signal is still at classical power levels before attenuation to the single-photon regime. Additionally, we propose placing a narrowband filter (say e.g. $0.05$~nm wide) centered on the reference wavelength, with all sidebands redirected to a dedicated detector: as we have shown, this allows for detection of spectral fluctuations in the output of the encoder laser that are generated by the manipulation of the \gls*{oil}.

Moreover, our results reveal that out-of-band reflections from the \gls*{oil} cavity create a clear pathway for high-power Trojan signals. Fortunately, this vulnerability can be effectively neutralized with straightforward measures. By limiting incoming light exclusively to the spectral range of the watchdog \gls*{pd}, simple and commercially available optical filters can readily suppress any unintended wavelengths. These devices introduce minimal additional optical loss, typically in the order of a few tenths of a decibel, and can be seamlessly integrated into existing setups without substantial redesign. 
This, coupled with a periodic spectral verification and monitoring of Alice and Bob's lasers, comprehensively addresses the identified vulnerabilities, significantly reinforcing the practical security of \gls*{oil}-based \gls*{tfqkd} systems without compromising their performance.

Although the proposed countermeasures will allow to close the identified security loopholes, we remark that their proper implementation is conditional on an accurate analysis of the system specifics and the application of an adequate security proof, establishing concrete values for the amount of required attenuation and filtering. To the best of our knowledge, to date no such proof exists, and while its derivation is beyond the scope of this paper, our results motivate the inclusion of information leakage induced by reference-beam attacks into future theoretical analyses of \gls*{tfqkd}.

More broadly, our analysis emphasizes the importance of restricting Eve’s control over the degrees of freedom of the injected optical signals, such as frequency, power, temporal shape, and polarization, to ensure robust security of quantum communication protocols. A rigorous and comprehensive restriction not only addresses currently identified attack vectors but also provides resilience against future, yet-undiscovered, attack strategies. Maintaining tight control over all optical parameters entering the encoder thus forms a foundational element for securing practical quantum communication systems in real-world deployments.

\section*{Data and materials availability}
All data are available from the corresponding author upon reasonable request.

\section*{Correspondence and requests for materials}
Should be addressed to S.J. (sergio.juarez@toshiba.eu) or A.M. (amarcomini@vqcc.uvigo.es).

\section*{Authors contributions}
S.J., A.M., R.I.W., and D.R. identified the academic motivation for this research project. S.J. and A.M. built the experimental setups, collected the measurements, and analyzed the results with support and supervision from M.P., R.I.W., T.D., R.M.S., and D.R., and particularly, derived the simulation results under the supervision of M.C.. S.J., A.M., and R.I.W. wrote the manuscript, with all authors contributing to its improvement and the verification of the results.

\section*{Acknowledgments}
The authors thank S. Morrissey, F. Gr\"unenfelder, O. Crampton, A. Brzosko, Y.S. Lo, G. Shooter, and P.R. Smith, for insightful discussions.
We acknowledge support from the European Union’s Horizon Europe Framework Programme under the Marie Sk\l{}odowska Curie Grant No.101072637, Project Quantum-Safe Internet (QSI). Views and opinions expressed are however those of the author(s) only and do not necessarily reflect those of the European Union.
Neither the European Union nor the granting authority can be held responsible for them.
We also acknowledge support from the Galician Regional Government (consolidation of research units: atlanTTic), the Spanish Ministry of Economy and Competitiveness (MINECO), the Fondo Europeo de Desarrollo Regional (FEDER) through the grant No. PID2024-162270OB-I00, MICIN with funding from the European Union NextGenerationEU (PRTR-C17.I1) and the Galician Regional Government with own funding through the “Planes Complementarios de I+D+I con las Comunidades Autonomas” in Quantum Communication, the “Hub Nacional de Excelencia en Comunicaciones Cuanticas” funded by the Spanish Ministry for Digital Transformation and the Public Service and the European Union NextGenerationEU, the European Union’s Horizon Europe Framework Programme under the project “Quantum Secure Networks Partnership” (QSNP, grant agreement No 101114043) and the European Union via the European Health and Digital Executive Agency (HADEA) under the Project QuTechSpace (grant 101135225).

\section*{Competing interests}
The authors declare no competing interests.

\small
\section*{Methods}

\subsection*{Experimental details of the fast intensity modulation attack}

In this section we present the details of our investigation on the modulation parameters for Eve, and discuss how for every configuration a sufficiently high repetition rate deceives slowly-integrating power meters, which therefore constitute insecure watchdogs.

\begin{figure}
    \centering
    \includegraphics[width=0.99\linewidth]{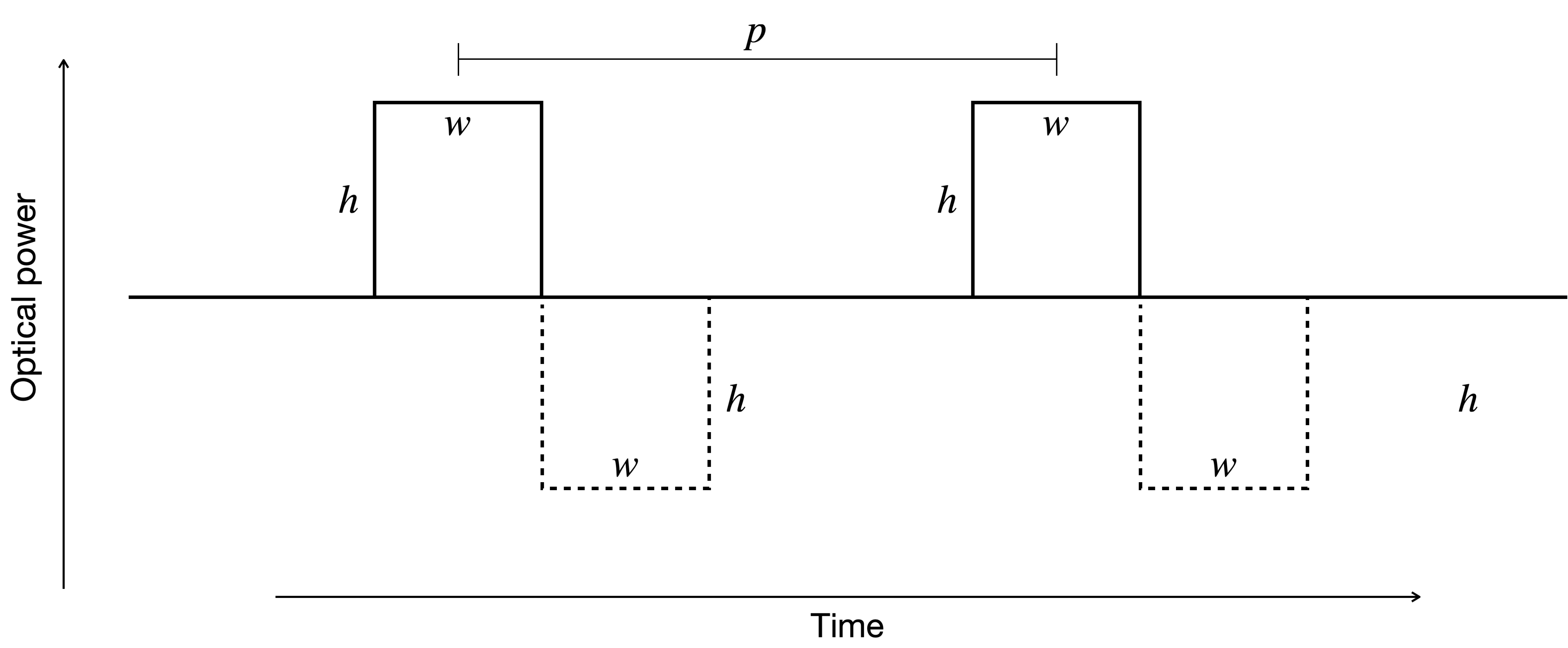}
    \caption{\textbf{Modulation example of Eve's attack}. The schematics describe the nominal input provided to the \gls*{Imod} in Fig.~\ref{fig:setups}. Solid lines refer to an ``Up" configuration where Eve periodically increases the input power, whereas by appending to this modulation a symmetric reduction in power to keep the average power constant (dashed lines), Eve can prepare the ``Up-to-Down" configuration. The actual modulated signal in the latter case is displayed in the first plot of Fig.~\ref{fig:Watchdog_results}(a). The ``Down" and ``Down-to-Up" configurations are obtained by considering vertical mirrored versions of these schemes. For each configuration we test different values of the pulse height $h$, the pulse width $w$ and the modulation period $p$.}
    \label{fig:Modulation_example}
\end{figure}

Fig.~\ref{fig:Modulation_example} reports an example of input pattern for Eve's intensity modulator for the case of a ``Up" modulation (that is, fast, periodic increases in the optical power of the reference signal), as well as for the ``Up-to-Down" modulation introduced in Fig.~\ref{fig:Watchdog_results}. The vertically mirrored patterns yield the ``Down" and the ``Down-to-Up" configurations, respectively. For each of the four configurations, we tested input and output statistics by changing the attack repetition rate (corresponding to $1/p$ from Fig.~\ref{fig:Modulation_example}) and the modulation width $w$ and height $h$. Crucially, as noted in Fig.~\ref{fig:Watchdog_results}, for both the ``Up-to-Down" and ``Down-to-Up" modulations of the reference signal, the locked light at Alice's output always results in a ``Down-to-Up" configuration. This phenomenon also occurs for the ``Up" and ``Down" schemes, but notably in these latter cases the induced peak above the baseline is much smaller than the inflection below the baseline, resulting in a worse quality of the injection and overall reduction of the mean photon density. As this will affect performance rather than security, here we focus on the analysis for the ``Up-to-Down" and ``Down-to-Up" configurations, which allow instead to induce a significant transient increase in the optical power.

For the repetition rate of the attack we tested values ranging from $100$MHz to $5$GHz, comparable with the operational rate of state-of-the-art \gls*{qkd} schemes. For all these options it holds $p \gg w$, meaning that the produced pulse shape is not altered, and no appreciable changes have been observed in the detection statistics. Therefore, in this analysis we consider a modulation rate of $1$GHz. As for the modulation width, we analyzed the produced pulse shapes while sweeping over the range $50$ps$-250$ps. We observed that a short modulation is preferable as it allows for a steeper transient between low and high levels of the photon density, which ultimately results in wider and taller peaks above the baseline. Therefore, results in this section refer to the setting choice $w=50$ps, corresponding to the minimal resolution of our controls. 

Finally, we tested various magnitudes of the modulation height $h$. Increasing this parameter directly impacts the modulation amplitude of the photon density. As this quantity is the most objective and meaningful metric for a device-agnostic analysis, figures in this paper show various experimental results as a function of Eve's normalized intensity modulation, corresponding to the peak-to-peak amplitude of the \gls*{snspd}s results displayed, for example, in Fig.~\ref{fig:Watchdog_results}(a). 

The full experimental results for the slow-integrating power meters are displayed in the \supp \ \cite{ThisPaperSM}, where figures report the normalized average power and normalized power fluctuations detected by the devices, respectively, and compare them with the attack-free scenario. 

The results discussed in the main text were obtained with an input power of 5~mW before the intensity modulator shown in Fig.~\ref{fig:setups}. 

Statistics are computed over $\mathcal{O}(10^5)$ samples for each configuration. We observe that regardless of the attack intensity and integration time, power meters fail to detect anomalies in the observed quantities, thus mistakenly assuming a safe scenario when Alice's laser is under a modulation attack. This motivates us to introduce novel surveillance procedures, namely the countermeasures introduced in the ``Experimental demonstration and countermeasures" section, so to guarantee the implementation security of \gls*{oil}-based \gls*{tfqkd} against this novel class of potential threats.

\subsection*{Simulation details}

Here we present the details of the \gls*{skr} simulations displayed in Fig.~\ref{fig:SKR}. Our analysis is based on the protocol description and security proof for the ``Sending-or-Not-Sending" (SNS) TF-QKD scheme introduced in Ref.~\cite{Wang.2018} and the simulations provided in Ref.~\cite{curty2019simple}. For simplicity, we consider the asymptotic case of infinitely many signals sent and infinitely many decoy intensities: the former implies that we can consider the key basis ($Z-$basis) to be chosen with probability $P_Z\to1$, while the second implies that all the required yields can be computed exactly. In addition, we consider a standard channel model, where losses are modeled as a beam splitter whose transmittance matches the one of the optical fiber.  

Let $\eta$ denote the total channel transmittance for a given distance between Alice and Bob and let $t=\sqrt{\eta}$ be the transmittance of the symmetric channels connecting each of the legitimate parties to Charlie. Here we consider a standard fiber loss coefficient $\alpha=0.2$ dB/km and detectors with dark count probability $p_d=10^{-8}$ and perfect detection efficiency (note that a finite detection efficiency could be incorporated in the definition of $\eta$). Following the definition of key events in Ref.~\cite{Wang.2018}, the vacuum and single-photon yields are found to be respectively \cite{Wang.2018,curty2019simple}
\begin{gather}
    s_0 = 2 p_d (1 - p_d), \\
    s_1 = 2 (1 - p_d) \left[p_d (1 - t) + \frac{t}{2} \right].
\end{gather}

Let $\mu$ denote the intensity of the coherent state in the $Z$ basis and $\epsilon$ the probability of Alice and Bob sending such coherent state (instead of vacuum) when they select a key round. For the total yield of correct and erroneous bits in the $Z-$basis, we have
\begin{equation}\label{eqn:S_z_corr}
    S_z^{\text{corr}} = 4 \epsilon (1 - \epsilon) (1 - p_d) e^{- \mu t} \left(p_d + e^{\mu t / 2} - 1\right),
\end{equation}
\begin{align}\nonumber
    S_z^{\text{err}} =& 2 \epsilon^2 (1 - p_d) e^{-2 \mu t} \left(p_d + e^{\mu t} I_0(\mu t \cos(\Delta\varphi)) - 1\right)\\
    &+ 2 (1 - \epsilon)^2 p_d (1 - p_d),
    \label{eqn:S_z_err}
\end{align}
where $\Delta\varphi$ quantifies the polarization misalignment and $I_0(x)$ denotes the modified Bessel function of the first kind and order zero. It follows that for the total $Z-$basis yield and \gls*{qber} we have:
\begin{equation}\label{eqn:SzEz}
    S_z = S_z^{\text{corr}} + S_z^{\text{err}} , \quad \quad \quad E_z = \frac{S_z^{\text{err}}}{S_z}.
\end{equation}

We remark that in the SNS-TF-QKD protocol the notion of a valid $X-$basis event, required to estimate the phase error rate using decoy states, is different from that of a $Z-$basis event. In detail, considering a decoy-state intensity $\nu$, we have that
\begin{equation}
S_x^{\text{corr}} = (1 - p_d) \left[e^{- \nu t (1 - \cos(\Delta\varphi) \cos(\Delta\theta))} - (1-p_d)e^{-2 \nu t} \right],
\end{equation}
\begin{equation}
S_x^{\text{err}} = (1 - p_d) \left[e^{- \nu t (1 + \cos(\Delta\varphi) \cos(\Delta\theta))} - (1-p_d)e^{-2 \nu t} \right],
\end{equation}
being $\Delta\theta$ the phase misalignment. The total $X-$basis yield and QBER follow as
\begin{equation}
    S_x = S_x^{\text{corr}} + S_x^{\text{err}}, \quad \quad \quad E_x = \frac{S_x^{\text{err}}}{S_x}.
\end{equation}
From these quantities, an upper bound on the phase error rate can be found to be \cite{Wang.2018}:
\begin{equation}
    e_{\text{ph}} \le \bar{e}_{\text{ph}} = \frac{S_x E_x - \frac{1}{2} s_0 e^{-2 \nu}}{2 \nu e^{-2 \nu} s_1}.
\end{equation}
The above bound holds for any value of $\nu$ and one can numerically check that it gets tighter as $\nu\ll1$. Therefore, in our calculations we considered the fixed value of $\nu=10^{-6}$.

Finally, the secret key rate $R$ can be obtain as
\begin{equation}\label{eqn: asymptotic SNS SKR}
    R = 2 \epsilon (1 - \epsilon) \mu e^{- \mu} s_1 [1 - h(\bar{e}_{\text{ph}})] - f_E \ S_z \ h(E_z),
\end{equation}
where $h(x)$ denotes the binary entropy of $x$, and $f_E=1.16$ is the error correction inefficiency.

As mentioned in the caption of Fig.~\ref{fig:SKR}, before \textit{actually} running the protocol Alice and Bob establish the optimal key parameters 
\begin{equation}
    (\tilde{\epsilon},  \tilde{\mu}) := \underset{\epsilon,\mu}{\text{argmax}} \ R
\end{equation}
for a given set of system parameters $\eta, p_d, \Delta\varphi$ and $\Delta\theta$. All curves in Fig.~\ref{fig:SKR} refer to the optimal case $\Delta\varphi = \Delta\theta = 0$, but no significant difference has been observed when comparing to higher misalignment cases.

Due to the attack, although Alice and Bob \textit{expect} to prepare $Z-$states with intensity $\tilde\mu$ they \textit{actually} prepare states with intensity $\tilde\mu'=\kappa\mu$, for some $\kappa>0$ ($\kappa=1.51$ in our case, from Fig.~\ref{fig:Watchdog_results}). With a noiseless channel model, the $Z-$basis statistics that Alice and Bob \textit{actually} observe are obtained through Eqs.~\ref{eqn:S_z_corr}, \ref{eqn:S_z_err} and \ref{eqn:SzEz} with $\mu=\tilde\mu'$. This shows that within this framework the attack has a direct impact on the $Z-$basis QBER, while its effect on the phase error rate is negligible as $\nu \approx 0$. \\
The \textit{actual} secret key rate guaranteed by the security proof in this case is obtained by plugging in the \textit{actual} intensity sent $\tilde\mu'$ and the \textit{actual} bit error rate $E_z$ in Eq.~\ref{eqn: asymptotic SNS SKR}, together with the sending probability $\tilde\epsilon$ which is unchanged.

Nevertheless, if Alice and Bob are unaware of the attack, they believe they are sending coherent states with the average intensity $\tilde\mu$ and observe the actual bit error rate $E_z$, as determined from Charlie’s announcements. As a result, they compute their expected key rate by substituting $\tilde\mu$ and $E_z$ into Eq.~\ref{eqn: asymptotic SNS SKR}, which leads to an erroneous overestimation.


\begin{thebibliography}{49}%
\makeatletter
\providecommand \@ifxundefined [1]{%
 \@ifx{#1\undefined}
}%
\providecommand \@ifnum [1]{%
 \ifnum #1\expandafter \@firstoftwo
 \else \expandafter \@secondoftwo
 \fi
}%
\providecommand \@ifx [1]{%
 \ifx #1\expandafter \@firstoftwo
 \else \expandafter \@secondoftwo
 \fi
}%
\providecommand \natexlab [1]{#1}%
\providecommand \enquote  [1]{``#1''}%
\providecommand \bibnamefont  [1]{#1}%
\providecommand \bibfnamefont [1]{#1}%
\providecommand \citenamefont [1]{#1}%
\providecommand \href@noop [0]{\@secondoftwo}%
\providecommand \href [0]{\begingroup \@sanitize@url \@href}%
\providecommand \@href[1]{\@@startlink{#1}\@@href}%
\providecommand \@@href[1]{\endgroup#1\@@endlink}%
\providecommand \@sanitize@url [0]{\catcode `\\12\catcode `\$12\catcode `\&12\catcode `\#12\catcode `\^12\catcode `\_12\catcode `\%12\relax}%
\providecommand \@@startlink[1]{}%
\providecommand \@@endlink[0]{}%
\providecommand \url  [0]{\begingroup\@sanitize@url \@url }%
\providecommand \@url [1]{\endgroup\@href {#1}{\urlprefix }}%
\providecommand \urlprefix  [0]{URL }%
\providecommand \Eprint [0]{\href }%
\providecommand \doibase [0]{https://doi.org/}%
\providecommand \selectlanguage [0]{\@gobble}%
\providecommand \bibinfo  [0]{\@secondoftwo}%
\providecommand \bibfield  [0]{\@secondoftwo}%
\providecommand \translation [1]{[#1]}%
\providecommand \BibitemOpen [0]{}%
\providecommand \bibitemStop [0]{}%
\providecommand \bibitemNoStop [0]{.\EOS\space}%
\providecommand \EOS [0]{\spacefactor3000\relax}%
\providecommand \BibitemShut  [1]{\csname bibitem#1\endcsname}%
\let\auto@bib@innerbib\@empty
\bibitem [{\citenamefont {Bennett}\ and\ \citenamefont {Brassard}(2014)}]{BB84}%
  \BibitemOpen
  \bibfield  {author} {\bibinfo {author} {\bibfnamefont {C.~H.}\ \bibnamefont {Bennett}}\ and\ \bibinfo {author} {\bibfnamefont {G.}~\bibnamefont {Brassard}},\ }\bibfield  {title} {\bibinfo {title} {Quantum cryptography: Public key distribution and coin tossing},\ }\href@noop {} {\bibfield  {journal} {\bibinfo  {journal} {Theoretical Computer Science}\ }\textbf {\bibinfo {volume} {560}},\ \bibinfo {pages} {7} (\bibinfo {year} {2014})}\BibitemShut {NoStop}%
\bibitem [{\citenamefont {Ekert}(1991)}]{Ekert.1991}%
  \BibitemOpen
  \bibfield  {author} {\bibinfo {author} {\bibfnamefont {A.~K.}\ \bibnamefont {Ekert}},\ }\bibfield  {title} {\bibinfo {title} {Quantum cryptography based on {B}ell's theorem},\ }\href@noop {} {\bibfield  {journal} {\bibinfo  {journal} {Physical Review Letters}\ }\textbf {\bibinfo {volume} {67}},\ \bibinfo {pages} {661} (\bibinfo {year} {1991})}\BibitemShut {NoStop}%
\bibitem [{\citenamefont {Pirandola}\ \emph {et~al.}(2020)\citenamefont {Pirandola}, \citenamefont {Andersen}, \citenamefont {Banchi}, \citenamefont {Berta}, \citenamefont {Bunandar}, \citenamefont {Colbeck}, \citenamefont {Englund}, \citenamefont {Gehring}, \citenamefont {Lupo}, \citenamefont {Ottaviani}, \citenamefont {Pereira}, \citenamefont {Razavi}, \citenamefont {{Shamsul Shaari}}, \citenamefont {Tomamichel}, \citenamefont {Usenko}, \citenamefont {Vallone}, \citenamefont {Villoresi},\ and\ \citenamefont {Wallden}}]{Pirandola.2020}%
  \BibitemOpen
  \bibfield  {author} {\bibinfo {author} {\bibfnamefont {S.}~\bibnamefont {Pirandola}}, \bibinfo {author} {\bibfnamefont {U.~L.}\ \bibnamefont {Andersen}}, \bibinfo {author} {\bibfnamefont {L.}~\bibnamefont {Banchi}}, \bibinfo {author} {\bibfnamefont {M.}~\bibnamefont {Berta}}, \bibinfo {author} {\bibfnamefont {D.}~\bibnamefont {Bunandar}}, \bibinfo {author} {\bibfnamefont {R.}~\bibnamefont {Colbeck}}, \bibinfo {author} {\bibfnamefont {D.}~\bibnamefont {Englund}}, \bibinfo {author} {\bibfnamefont {T.}~\bibnamefont {Gehring}}, \bibinfo {author} {\bibfnamefont {C.}~\bibnamefont {Lupo}}, \bibinfo {author} {\bibfnamefont {C.}~\bibnamefont {Ottaviani}}, \bibinfo {author} {\bibfnamefont {J.~L.}\ \bibnamefont {Pereira}}, \bibinfo {author} {\bibfnamefont {M.}~\bibnamefont {Razavi}}, \bibinfo {author} {\bibfnamefont {J.}~\bibnamefont {{Shamsul Shaari}}}, \bibinfo {author} {\bibfnamefont {M.}~\bibnamefont {Tomamichel}}, \bibinfo {author} {\bibfnamefont {V.~C.}\ \bibnamefont {Usenko}}, \bibinfo {author} {\bibfnamefont
  {G.}~\bibnamefont {Vallone}}, \bibinfo {author} {\bibfnamefont {P.}~\bibnamefont {Villoresi}},\ and\ \bibinfo {author} {\bibfnamefont {P.}~\bibnamefont {Wallden}},\ }\bibfield  {title} {\bibinfo {title} {Advances in quantum cryptography},\ }\href@noop {} {\bibfield  {journal} {\bibinfo  {journal} {Advances in Optics and Photonics}\ }\textbf {\bibinfo {volume} {12}},\ \bibinfo {pages} {1012} (\bibinfo {year} {2020})}\BibitemShut {NoStop}%
\bibitem [{\citenamefont {Lo}\ \emph {et~al.}(2014)\citenamefont {Lo}, \citenamefont {Curty},\ and\ \citenamefont {Tamaki}}]{lo2014secure}%
  \BibitemOpen
  \bibfield  {author} {\bibinfo {author} {\bibfnamefont {H.-K.}\ \bibnamefont {Lo}}, \bibinfo {author} {\bibfnamefont {M.}~\bibnamefont {Curty}},\ and\ \bibinfo {author} {\bibfnamefont {K.}~\bibnamefont {Tamaki}},\ }\bibfield  {title} {\bibinfo {title} {Secure quantum key distribution},\ }\href@noop {} {\bibfield  {journal} {\bibinfo  {journal} {Nature Photonics}\ }\textbf {\bibinfo {volume} {8}},\ \bibinfo {pages} {595} (\bibinfo {year} {2014})}\BibitemShut {NoStop}%
\bibitem [{\citenamefont {Sasaki}\ \emph {et~al.}(2011)\citenamefont {Sasaki}, \citenamefont {Fujiwara}, \citenamefont {Ishizuka}, \citenamefont {Klaus}, \citenamefont {Wakui}, \citenamefont {Takeoka}, \citenamefont {Miki}, \citenamefont {Yamashita}, \citenamefont {Wang}, \citenamefont {Tanaka} \emph {et~al.}}]{sasaki2011field}%
  \BibitemOpen
  \bibfield  {author} {\bibinfo {author} {\bibfnamefont {M.}~\bibnamefont {Sasaki}}, \bibinfo {author} {\bibfnamefont {M.}~\bibnamefont {Fujiwara}}, \bibinfo {author} {\bibfnamefont {H.}~\bibnamefont {Ishizuka}}, \bibinfo {author} {\bibfnamefont {W.}~\bibnamefont {Klaus}}, \bibinfo {author} {\bibfnamefont {K.}~\bibnamefont {Wakui}}, \bibinfo {author} {\bibfnamefont {M.}~\bibnamefont {Takeoka}}, \bibinfo {author} {\bibfnamefont {S.}~\bibnamefont {Miki}}, \bibinfo {author} {\bibfnamefont {T.}~\bibnamefont {Yamashita}}, \bibinfo {author} {\bibfnamefont {Z.}~\bibnamefont {Wang}}, \bibinfo {author} {\bibfnamefont {A.}~\bibnamefont {Tanaka}}, \emph {et~al.},\ }\bibfield  {title} {\bibinfo {title} {Field test of quantum key distribution in the tokyo qkd network},\ }\href@noop {} {\bibfield  {journal} {\bibinfo  {journal} {Optics express}\ }\textbf {\bibinfo {volume} {19}},\ \bibinfo {pages} {10387} (\bibinfo {year} {2011})}\BibitemShut {NoStop}%
\bibitem [{\citenamefont {Martin}\ \emph {et~al.}(2024)\citenamefont {Martin}, \citenamefont {Brito}, \citenamefont {Ort{\'\i}z}, \citenamefont {M{\'e}ndez}, \citenamefont {Buruaga}, \citenamefont {Vicente}, \citenamefont {Sebastian-Lombrana}, \citenamefont {Rincon}, \citenamefont {Perez}, \citenamefont {Sanchez} \emph {et~al.}}]{martin2024madqci}%
  \BibitemOpen
  \bibfield  {author} {\bibinfo {author} {\bibfnamefont {V.}~\bibnamefont {Martin}}, \bibinfo {author} {\bibfnamefont {J.~P.}\ \bibnamefont {Brito}}, \bibinfo {author} {\bibfnamefont {L.}~\bibnamefont {Ort{\'\i}z}}, \bibinfo {author} {\bibfnamefont {R.}~\bibnamefont {M{\'e}ndez}}, \bibinfo {author} {\bibfnamefont {J.}~\bibnamefont {Buruaga}}, \bibinfo {author} {\bibfnamefont {R.}~\bibnamefont {Vicente}}, \bibinfo {author} {\bibfnamefont {A.}~\bibnamefont {Sebastian-Lombrana}}, \bibinfo {author} {\bibfnamefont {D.}~\bibnamefont {Rincon}}, \bibinfo {author} {\bibfnamefont {F.}~\bibnamefont {Perez}}, \bibinfo {author} {\bibfnamefont {C.}~\bibnamefont {Sanchez}}, \emph {et~al.},\ }\bibfield  {title} {\bibinfo {title} {Madqci: a heterogeneous and scalable sdn-qkd network deployed in production facilities},\ }\href@noop {} {\bibfield  {journal} {\bibinfo  {journal} {npj Quantum Information}\ }\textbf {\bibinfo {volume} {10}},\ \bibinfo {pages} {80} (\bibinfo {year} {2024})}\BibitemShut {NoStop}%
\bibitem [{\citenamefont {Chen}\ \emph {et~al.}(2025)\citenamefont {Chen}, \citenamefont {Li}, \citenamefont {Wang}, \citenamefont {Zhao}, \citenamefont {Ye}, \citenamefont {Li}, \citenamefont {Chen}, \citenamefont {Han}, \citenamefont {Tang}, \citenamefont {Miao} \emph {et~al.}}]{chen2025implementation}%
  \BibitemOpen
  \bibfield  {author} {\bibinfo {author} {\bibfnamefont {H.-Z.}\ \bibnamefont {Chen}}, \bibinfo {author} {\bibfnamefont {M.-H.}\ \bibnamefont {Li}}, \bibinfo {author} {\bibfnamefont {Y.~Z.}\ \bibnamefont {Wang}}, \bibinfo {author} {\bibfnamefont {Z.-G.}\ \bibnamefont {Zhao}}, \bibinfo {author} {\bibfnamefont {C.}~\bibnamefont {Ye}}, \bibinfo {author} {\bibfnamefont {F.~L.}\ \bibnamefont {Li}}, \bibinfo {author} {\bibfnamefont {Z.}~\bibnamefont {Chen}}, \bibinfo {author} {\bibfnamefont {S.-L.}\ \bibnamefont {Han}}, \bibinfo {author} {\bibfnamefont {B.}~\bibnamefont {Tang}}, \bibinfo {author} {\bibfnamefont {Y.~J.}\ \bibnamefont {Miao}}, \emph {et~al.},\ }\bibfield  {title} {\bibinfo {title} {Implementation of carrier-grade quantum communication networks over 10000 km},\ }\href@noop {} {\bibfield  {journal} {\bibinfo  {journal} {npj Quantum Information}\ }\textbf {\bibinfo {volume} {11}},\ \bibinfo {pages} {137} (\bibinfo {year} {2025})}\BibitemShut {NoStop}%
\bibitem [{\citenamefont {Lucamarini}\ \emph {et~al.}(2018)\citenamefont {Lucamarini}, \citenamefont {Yuan}, \citenamefont {Dynes},\ and\ \citenamefont {Shields}}]{Lucamarini.2018}%
  \BibitemOpen
  \bibfield  {author} {\bibinfo {author} {\bibfnamefont {M.}~\bibnamefont {Lucamarini}}, \bibinfo {author} {\bibfnamefont {Z.~L.}\ \bibnamefont {Yuan}}, \bibinfo {author} {\bibfnamefont {J.~F.}\ \bibnamefont {Dynes}},\ and\ \bibinfo {author} {\bibfnamefont {A.~J.}\ \bibnamefont {Shields}},\ }\bibfield  {title} {\bibinfo {title} {Overcoming the rate-distance limit of quantum key distribution without quantum repeaters},\ }\href@noop {} {\bibfield  {journal} {\bibinfo  {journal} {Nature}\ }\textbf {\bibinfo {volume} {557}},\ \bibinfo {pages} {400} (\bibinfo {year} {2018})}\BibitemShut {NoStop}%
\bibitem [{\citenamefont {Wang}\ \emph {et~al.}(2018{\natexlab{a}})\citenamefont {Wang}, \citenamefont {Yu},\ and\ \citenamefont {Hu}}]{Wang.2018}%
  \BibitemOpen
  \bibfield  {author} {\bibinfo {author} {\bibfnamefont {X.-B.}\ \bibnamefont {Wang}}, \bibinfo {author} {\bibfnamefont {Z.-W.}\ \bibnamefont {Yu}},\ and\ \bibinfo {author} {\bibfnamefont {X.-L.}\ \bibnamefont {Hu}},\ }\bibfield  {title} {\bibinfo {title} {Twin-field quantum key distribution with large misalignment error},\ }\href@noop {} {\bibfield  {journal} {\bibinfo  {journal} {Physical Review A}\ }\textbf {\bibinfo {volume} {98}} (\bibinfo {year} {2018}{\natexlab{a}})}\BibitemShut {NoStop}%
\bibitem [{\citenamefont {Xu}\ \emph {et~al.}(2020{\natexlab{a}})\citenamefont {Xu}, \citenamefont {Yu}, \citenamefont {Jiang}, \citenamefont {Hu},\ and\ \citenamefont {Wang}}]{Xu.2020}%
  \BibitemOpen
  \bibfield  {author} {\bibinfo {author} {\bibfnamefont {H.}~\bibnamefont {Xu}}, \bibinfo {author} {\bibfnamefont {Z.-W.}\ \bibnamefont {Yu}}, \bibinfo {author} {\bibfnamefont {C.}~\bibnamefont {Jiang}}, \bibinfo {author} {\bibfnamefont {X.-L.}\ \bibnamefont {Hu}},\ and\ \bibinfo {author} {\bibfnamefont {X.-B.}\ \bibnamefont {Wang}},\ }\bibfield  {title} {\bibinfo {title} {Sending-or-not-sending twin-field quantum key distribution: Breaking the direct transmission key rate},\ }\href@noop {} {\bibfield  {journal} {\bibinfo  {journal} {Physical Review A}\ }\textbf {\bibinfo {volume} {101}} (\bibinfo {year} {2020}{\natexlab{a}})}\BibitemShut {NoStop}%
\bibitem [{\citenamefont {Curty}\ \emph {et~al.}(2019)\citenamefont {Curty}, \citenamefont {Azuma},\ and\ \citenamefont {Lo}}]{curty2019simple}%
  \BibitemOpen
  \bibfield  {author} {\bibinfo {author} {\bibfnamefont {M.}~\bibnamefont {Curty}}, \bibinfo {author} {\bibfnamefont {K.}~\bibnamefont {Azuma}},\ and\ \bibinfo {author} {\bibfnamefont {H.-K.}\ \bibnamefont {Lo}},\ }\bibfield  {title} {\bibinfo {title} {Simple security proof of twin-field type quantum key distribution protocol},\ }\href@noop {} {\bibfield  {journal} {\bibinfo  {journal} {npj Quantum Information}\ }\textbf {\bibinfo {volume} {5}},\ \bibinfo {pages} {64} (\bibinfo {year} {2019})}\BibitemShut {NoStop}%
\bibitem [{\citenamefont {Curty}\ \emph {et~al.}(2009)\citenamefont {Curty}, \citenamefont {Moroder}, \citenamefont {Ma}, \citenamefont {Lo},\ and\ \citenamefont {L\"utkenhaus}}]{curty.2009}%
  \BibitemOpen
  \bibfield  {author} {\bibinfo {author} {\bibfnamefont {M.}~\bibnamefont {Curty}}, \bibinfo {author} {\bibfnamefont {T.}~\bibnamefont {Moroder}}, \bibinfo {author} {\bibfnamefont {X.}~\bibnamefont {Ma}}, \bibinfo {author} {\bibfnamefont {H.-K.}\ \bibnamefont {Lo}},\ and\ \bibinfo {author} {\bibfnamefont {N.}~\bibnamefont {L\"utkenhaus}},\ }\bibfield  {title} {\bibinfo {title} {Upper bounds for the secure key rate of the decoy-state quantum key distribution},\ }\href@noop {} {\bibfield  {journal} {\bibinfo  {journal} {Phys. Rev. A}\ }\textbf {\bibinfo {volume} {79}},\ \bibinfo {pages} {032335} (\bibinfo {year} {2009})}\BibitemShut {NoStop}%
\bibitem [{\citenamefont {Pirandola}\ \emph {et~al.}(2017)\citenamefont {Pirandola}, \citenamefont {Laurenza}, \citenamefont {Ottaviani},\ and\ \citenamefont {Banchi}}]{Pirandola.2017}%
  \BibitemOpen
  \bibfield  {author} {\bibinfo {author} {\bibfnamefont {S.}~\bibnamefont {Pirandola}}, \bibinfo {author} {\bibfnamefont {R.}~\bibnamefont {Laurenza}}, \bibinfo {author} {\bibfnamefont {C.}~\bibnamefont {Ottaviani}},\ and\ \bibinfo {author} {\bibfnamefont {L.}~\bibnamefont {Banchi}},\ }\bibfield  {title} {\bibinfo {title} {Fundamental limits of repeaterless quantum communications},\ }\href@noop {} {\bibfield  {journal} {\bibinfo  {journal} {Nature Communications}\ }\textbf {\bibinfo {volume} {8}},\ \bibinfo {pages} {15043} (\bibinfo {year} {2017})}\BibitemShut {NoStop}%
\bibitem [{\citenamefont {Minder}\ \emph {et~al.}(2019)\citenamefont {Minder}, \citenamefont {Pittaluga}, \citenamefont {Roberts}, \citenamefont {Lucamarini}, \citenamefont {Dynes}, \citenamefont {Yuan},\ and\ \citenamefont {Shields}}]{Minder.2019}%
  \BibitemOpen
  \bibfield  {author} {\bibinfo {author} {\bibfnamefont {M.}~\bibnamefont {Minder}}, \bibinfo {author} {\bibfnamefont {M.}~\bibnamefont {Pittaluga}}, \bibinfo {author} {\bibfnamefont {G.~L.}\ \bibnamefont {Roberts}}, \bibinfo {author} {\bibfnamefont {M.}~\bibnamefont {Lucamarini}}, \bibinfo {author} {\bibfnamefont {J.~F.}\ \bibnamefont {Dynes}}, \bibinfo {author} {\bibfnamefont {Z.~L.}\ \bibnamefont {Yuan}},\ and\ \bibinfo {author} {\bibfnamefont {A.~J.}\ \bibnamefont {Shields}},\ }\bibfield  {title} {\bibinfo {title} {Experimental quantum key distribution beyond the repeaterless secret key capacity},\ }\href@noop {} {\bibfield  {journal} {\bibinfo  {journal} {Nature Photonics}\ }\textbf {\bibinfo {volume} {13}},\ \bibinfo {pages} {334} (\bibinfo {year} {2019})}\BibitemShut {NoStop}%
\bibitem [{\citenamefont {Pittaluga}\ \emph {et~al.}(2021)\citenamefont {Pittaluga}, \citenamefont {Minder}, \citenamefont {Lucamarini}, \citenamefont {Sanzaro}, \citenamefont {Woodward}, \citenamefont {Li}, \citenamefont {Yuan},\ and\ \citenamefont {Shields}}]{Pittaluga.2021}%
  \BibitemOpen
  \bibfield  {author} {\bibinfo {author} {\bibfnamefont {M.}~\bibnamefont {Pittaluga}}, \bibinfo {author} {\bibfnamefont {M.}~\bibnamefont {Minder}}, \bibinfo {author} {\bibfnamefont {M.}~\bibnamefont {Lucamarini}}, \bibinfo {author} {\bibfnamefont {M.}~\bibnamefont {Sanzaro}}, \bibinfo {author} {\bibfnamefont {R.~I.}\ \bibnamefont {Woodward}}, \bibinfo {author} {\bibfnamefont {M.-J.}\ \bibnamefont {Li}}, \bibinfo {author} {\bibfnamefont {Z.}~\bibnamefont {Yuan}},\ and\ \bibinfo {author} {\bibfnamefont {A.~J.}\ \bibnamefont {Shields}},\ }\bibfield  {title} {\bibinfo {title} {600-km repeater-like quantum communications with dual-band stabilization},\ }\href@noop {} {\bibfield  {journal} {\bibinfo  {journal} {Nature Photonics}\ }\textbf {\bibinfo {volume} {15}},\ \bibinfo {pages} {530} (\bibinfo {year} {2021})}\BibitemShut {NoStop}%
\bibitem [{\citenamefont {Chen}\ \emph {et~al.}(2022)\citenamefont {Chen}, \citenamefont {Zhang}, \citenamefont {Liu}, \citenamefont {Jiang}, \citenamefont {Zhao}, \citenamefont {Zhang}, \citenamefont {Chen}, \citenamefont {Li}, \citenamefont {You}, \citenamefont {Wang}, \citenamefont {Chen}, \citenamefont {Wang}, \citenamefont {Zhang},\ and\ \citenamefont {Pan}}]{Chen.2022}%
  \BibitemOpen
  \bibfield  {author} {\bibinfo {author} {\bibfnamefont {J.-P.}\ \bibnamefont {Chen}}, \bibinfo {author} {\bibfnamefont {C.}~\bibnamefont {Zhang}}, \bibinfo {author} {\bibfnamefont {Y.}~\bibnamefont {Liu}}, \bibinfo {author} {\bibfnamefont {C.}~\bibnamefont {Jiang}}, \bibinfo {author} {\bibfnamefont {D.-F.}\ \bibnamefont {Zhao}}, \bibinfo {author} {\bibfnamefont {W.-J.}\ \bibnamefont {Zhang}}, \bibinfo {author} {\bibfnamefont {F.-X.}\ \bibnamefont {Chen}}, \bibinfo {author} {\bibfnamefont {H.}~\bibnamefont {Li}}, \bibinfo {author} {\bibfnamefont {L.-X.}\ \bibnamefont {You}}, \bibinfo {author} {\bibfnamefont {Z.}~\bibnamefont {Wang}}, \bibinfo {author} {\bibfnamefont {Y.}~\bibnamefont {Chen}}, \bibinfo {author} {\bibfnamefont {X.-B.}\ \bibnamefont {Wang}}, \bibinfo {author} {\bibfnamefont {Q.}~\bibnamefont {Zhang}},\ and\ \bibinfo {author} {\bibfnamefont {J.-W.}\ \bibnamefont {Pan}},\ }\bibfield  {title} {\bibinfo {title} {Quantum key distribution over 658 km fiber with distributed vibration sensing},\
  }\href@noop {} {\bibfield  {journal} {\bibinfo  {journal} {Physical Review Letters}\ }\textbf {\bibinfo {volume} {128}},\ \bibinfo {pages} {180502} (\bibinfo {year} {2022})}\BibitemShut {NoStop}%
\bibitem [{\citenamefont {Wang}\ \emph {et~al.}(2022)\citenamefont {Wang}, \citenamefont {Yin}, \citenamefont {He}, \citenamefont {Chen}, \citenamefont {Wang}, \citenamefont {Ye}, \citenamefont {Zhou}, \citenamefont {Fan-Yuan}, \citenamefont {Wang}, \citenamefont {Zhu}, \citenamefont {Morozov}, \citenamefont {Divochiy}, \citenamefont {Zhou}, \citenamefont {Guo},\ and\ \citenamefont {Han}}]{Wang.2022}%
  \BibitemOpen
  \bibfield  {author} {\bibinfo {author} {\bibfnamefont {S.}~\bibnamefont {Wang}}, \bibinfo {author} {\bibfnamefont {Z.-Q.}\ \bibnamefont {Yin}}, \bibinfo {author} {\bibfnamefont {D.-Y.}\ \bibnamefont {He}}, \bibinfo {author} {\bibfnamefont {W.}~\bibnamefont {Chen}}, \bibinfo {author} {\bibfnamefont {R.-Q.}\ \bibnamefont {Wang}}, \bibinfo {author} {\bibfnamefont {P.}~\bibnamefont {Ye}}, \bibinfo {author} {\bibfnamefont {Y.}~\bibnamefont {Zhou}}, \bibinfo {author} {\bibfnamefont {G.-J.}\ \bibnamefont {Fan-Yuan}}, \bibinfo {author} {\bibfnamefont {F.-X.}\ \bibnamefont {Wang}}, \bibinfo {author} {\bibfnamefont {Y.-G.}\ \bibnamefont {Zhu}}, \bibinfo {author} {\bibfnamefont {P.~V.}\ \bibnamefont {Morozov}}, \bibinfo {author} {\bibfnamefont {A.~V.}\ \bibnamefont {Divochiy}}, \bibinfo {author} {\bibfnamefont {Z.}~\bibnamefont {Zhou}}, \bibinfo {author} {\bibfnamefont {G.-C.}\ \bibnamefont {Guo}},\ and\ \bibinfo {author} {\bibfnamefont {Z.-F.}\ \bibnamefont {Han}},\ }\bibfield  {title} {\bibinfo {title} {Twin-field
  quantum key distribution over 830-km fibre},\ }\href@noop {} {\bibfield  {journal} {\bibinfo  {journal} {Nature Photonics}\ }\textbf {\bibinfo {volume} {16}},\ \bibinfo {pages} {154} (\bibinfo {year} {2022})}\BibitemShut {NoStop}%
\bibitem [{\citenamefont {Liu}\ \emph {et~al.}(2023)\citenamefont {Liu}, \citenamefont {Zhang}, \citenamefont {Jiang}, \citenamefont {Chen}, \citenamefont {Zhang}, \citenamefont {Pan}, \citenamefont {{Di Ma}}, \citenamefont {Dong}, \citenamefont {Xiong}, \citenamefont {Zhang}, \citenamefont {Li}, \citenamefont {Wang}, \citenamefont {Wu}, \citenamefont {Chen}, \citenamefont {You}, \citenamefont {Wang}, \citenamefont {Zhang},\ and\ \citenamefont {Pan}}]{Liu.2023}%
  \BibitemOpen
  \bibfield  {author} {\bibinfo {author} {\bibfnamefont {Y.}~\bibnamefont {Liu}}, \bibinfo {author} {\bibfnamefont {W.-J.}\ \bibnamefont {Zhang}}, \bibinfo {author} {\bibfnamefont {C.}~\bibnamefont {Jiang}}, \bibinfo {author} {\bibfnamefont {J.-P.}\ \bibnamefont {Chen}}, \bibinfo {author} {\bibfnamefont {C.}~\bibnamefont {Zhang}}, \bibinfo {author} {\bibfnamefont {W.-X.}\ \bibnamefont {Pan}}, \bibinfo {author} {\bibnamefont {{Di Ma}}}, \bibinfo {author} {\bibfnamefont {H.}~\bibnamefont {Dong}}, \bibinfo {author} {\bibfnamefont {J.-M.}\ \bibnamefont {Xiong}}, \bibinfo {author} {\bibfnamefont {C.-J.}\ \bibnamefont {Zhang}}, \bibinfo {author} {\bibfnamefont {H.}~\bibnamefont {Li}}, \bibinfo {author} {\bibfnamefont {R.-C.}\ \bibnamefont {Wang}}, \bibinfo {author} {\bibfnamefont {J.}~\bibnamefont {Wu}}, \bibinfo {author} {\bibfnamefont {T.-Y.}\ \bibnamefont {Chen}}, \bibinfo {author} {\bibfnamefont {L.}~\bibnamefont {You}}, \bibinfo {author} {\bibfnamefont {X.-B.}\ \bibnamefont {Wang}}, \bibinfo {author}
  {\bibfnamefont {Q.}~\bibnamefont {Zhang}},\ and\ \bibinfo {author} {\bibfnamefont {J.-W.}\ \bibnamefont {Pan}},\ }\bibfield  {title} {\bibinfo {title} {Experimental twin-field quantum key distribution over 1000 km fiber distance},\ }\href@noop {} {\bibfield  {journal} {\bibinfo  {journal} {Physical Review Letters}\ }\textbf {\bibinfo {volume} {130}},\ \bibinfo {pages} {210801} (\bibinfo {year} {2023})}\BibitemShut {NoStop}%
\bibitem [{\citenamefont {Pittaluga}\ \emph {et~al.}(2025)\citenamefont {Pittaluga}, \citenamefont {Lo}, \citenamefont {Brzosko}, \citenamefont {Woodward}, \citenamefont {Scalcon}, \citenamefont {Winnel}, \citenamefont {Roger}, \citenamefont {Dynes}, \citenamefont {Owen}, \citenamefont {Ju{\'a}rez} \emph {et~al.}}]{pittaluga2025long}%
  \BibitemOpen
  \bibfield  {author} {\bibinfo {author} {\bibfnamefont {M.}~\bibnamefont {Pittaluga}}, \bibinfo {author} {\bibfnamefont {Y.~S.}\ \bibnamefont {Lo}}, \bibinfo {author} {\bibfnamefont {A.}~\bibnamefont {Brzosko}}, \bibinfo {author} {\bibfnamefont {R.~I.}\ \bibnamefont {Woodward}}, \bibinfo {author} {\bibfnamefont {D.}~\bibnamefont {Scalcon}}, \bibinfo {author} {\bibfnamefont {M.~S.}\ \bibnamefont {Winnel}}, \bibinfo {author} {\bibfnamefont {T.}~\bibnamefont {Roger}}, \bibinfo {author} {\bibfnamefont {J.~F.}\ \bibnamefont {Dynes}}, \bibinfo {author} {\bibfnamefont {K.~A.}\ \bibnamefont {Owen}}, \bibinfo {author} {\bibfnamefont {S.}~\bibnamefont {Ju{\'a}rez}}, \emph {et~al.},\ }\bibfield  {title} {\bibinfo {title} {Long-distance coherent quantum communications in deployed telecom networks},\ }\href@noop {} {\bibfield  {journal} {\bibinfo  {journal} {Nature}\ }\textbf {\bibinfo {volume} {640}},\ \bibinfo {pages} {911} (\bibinfo {year} {2025})}\BibitemShut {NoStop}%
\bibitem [{\citenamefont {Chen}\ \emph {et~al.}(2021)\citenamefont {Chen}, \citenamefont {Zhang}, \citenamefont {Liu}, \citenamefont {Jiang}, \citenamefont {Zhang}, \citenamefont {Han}, \citenamefont {Ma}, \citenamefont {Hu}, \citenamefont {Li}, \citenamefont {Liu}, \citenamefont {Zhou}, \citenamefont {Jiang}, \citenamefont {Chen}, \citenamefont {Li}, \citenamefont {You}, \citenamefont {Wang}, \citenamefont {Wang}, \citenamefont {Zhang},\ and\ \citenamefont {Pan}}]{Chen.2021}%
  \BibitemOpen
  \bibfield  {author} {\bibinfo {author} {\bibfnamefont {J.-P.}\ \bibnamefont {Chen}}, \bibinfo {author} {\bibfnamefont {C.}~\bibnamefont {Zhang}}, \bibinfo {author} {\bibfnamefont {Y.}~\bibnamefont {Liu}}, \bibinfo {author} {\bibfnamefont {C.}~\bibnamefont {Jiang}}, \bibinfo {author} {\bibfnamefont {W.-J.}\ \bibnamefont {Zhang}}, \bibinfo {author} {\bibfnamefont {Z.-Y.}\ \bibnamefont {Han}}, \bibinfo {author} {\bibfnamefont {S.-Z.}\ \bibnamefont {Ma}}, \bibinfo {author} {\bibfnamefont {X.-L.}\ \bibnamefont {Hu}}, \bibinfo {author} {\bibfnamefont {Y.-H.}\ \bibnamefont {Li}}, \bibinfo {author} {\bibfnamefont {H.}~\bibnamefont {Liu}}, \bibinfo {author} {\bibfnamefont {F.}~\bibnamefont {Zhou}}, \bibinfo {author} {\bibfnamefont {H.-F.}\ \bibnamefont {Jiang}}, \bibinfo {author} {\bibfnamefont {T.-Y.}\ \bibnamefont {Chen}}, \bibinfo {author} {\bibfnamefont {H.}~\bibnamefont {Li}}, \bibinfo {author} {\bibfnamefont {L.-X.}\ \bibnamefont {You}}, \bibinfo {author} {\bibfnamefont {Z.}~\bibnamefont {Wang}}, \bibinfo
  {author} {\bibfnamefont {X.-B.}\ \bibnamefont {Wang}}, \bibinfo {author} {\bibfnamefont {Q.}~\bibnamefont {Zhang}},\ and\ \bibinfo {author} {\bibfnamefont {J.-W.}\ \bibnamefont {Pan}},\ }\bibfield  {title} {\bibinfo {title} {Twin-field quantum key distribution over a 511 km optical fibre linking two distant metropolitan areas},\ }\href@noop {} {\bibfield  {journal} {\bibinfo  {journal} {Nature Photonics}\ }\textbf {\bibinfo {volume} {299}},\ \bibinfo {pages} {1476} (\bibinfo {year} {2021})}\BibitemShut {NoStop}%
\bibitem [{\citenamefont {Liu}\ \emph {et~al.}(2021)\citenamefont {Liu}, \citenamefont {Jiang}, \citenamefont {Zhu}, \citenamefont {Zou}, \citenamefont {Yu}, \citenamefont {Hu}, \citenamefont {Xu}, \citenamefont {Ma}, \citenamefont {Han}, \citenamefont {Chen}, \citenamefont {Dai}, \citenamefont {Tang}, \citenamefont {Zhang}, \citenamefont {Li}, \citenamefont {You}, \citenamefont {Wang}, \citenamefont {Hua}, \citenamefont {Hu}, \citenamefont {Zhang}, \citenamefont {Zhou}, \citenamefont {Zhang}, \citenamefont {Wang}, \citenamefont {Chen},\ and\ \citenamefont {Pan}}]{Liu.2021}%
  \BibitemOpen
  \bibfield  {author} {\bibinfo {author} {\bibfnamefont {H.}~\bibnamefont {Liu}}, \bibinfo {author} {\bibfnamefont {C.}~\bibnamefont {Jiang}}, \bibinfo {author} {\bibfnamefont {H.-T.}\ \bibnamefont {Zhu}}, \bibinfo {author} {\bibfnamefont {M.}~\bibnamefont {Zou}}, \bibinfo {author} {\bibfnamefont {Z.-W.}\ \bibnamefont {Yu}}, \bibinfo {author} {\bibfnamefont {X.-L.}\ \bibnamefont {Hu}}, \bibinfo {author} {\bibfnamefont {H.}~\bibnamefont {Xu}}, \bibinfo {author} {\bibfnamefont {S.}~\bibnamefont {Ma}}, \bibinfo {author} {\bibfnamefont {Z.}~\bibnamefont {Han}}, \bibinfo {author} {\bibfnamefont {J.-P.}\ \bibnamefont {Chen}}, \bibinfo {author} {\bibfnamefont {Y.}~\bibnamefont {Dai}}, \bibinfo {author} {\bibfnamefont {S.-B.}\ \bibnamefont {Tang}}, \bibinfo {author} {\bibfnamefont {W.}~\bibnamefont {Zhang}}, \bibinfo {author} {\bibfnamefont {H.}~\bibnamefont {Li}}, \bibinfo {author} {\bibfnamefont {L.}~\bibnamefont {You}}, \bibinfo {author} {\bibfnamefont {Z.}~\bibnamefont {Wang}}, \bibinfo {author} {\bibfnamefont
  {Y.}~\bibnamefont {Hua}}, \bibinfo {author} {\bibfnamefont {H.}~\bibnamefont {Hu}}, \bibinfo {author} {\bibfnamefont {H.}~\bibnamefont {Zhang}}, \bibinfo {author} {\bibfnamefont {F.}~\bibnamefont {Zhou}}, \bibinfo {author} {\bibfnamefont {Q.}~\bibnamefont {Zhang}}, \bibinfo {author} {\bibfnamefont {X.-B.}\ \bibnamefont {Wang}}, \bibinfo {author} {\bibfnamefont {T.-Y.}\ \bibnamefont {Chen}},\ and\ \bibinfo {author} {\bibfnamefont {J.-W.}\ \bibnamefont {Pan}},\ }\bibfield  {title} {\bibinfo {title} {Field test of twin-field quantum key distribution through sending-or-not-sending over 428 km},\ }\href@noop {} {\bibfield  {journal} {\bibinfo  {journal} {Physical Review Letters}\ }\textbf {\bibinfo {volume} {126}} (\bibinfo {year} {2021})}\BibitemShut {NoStop}%
\bibitem [{\citenamefont {Clivati}\ \emph {et~al.}(2022)\citenamefont {Clivati}, \citenamefont {Meda}, \citenamefont {Donadello}, \citenamefont {Virz{\`i}}, \citenamefont {Genovese}, \citenamefont {Levi}, \citenamefont {Mura}, \citenamefont {Pittaluga}, \citenamefont {Yuan}, \citenamefont {Shields}, \citenamefont {Lucamarini}, \citenamefont {Degiovanni},\ and\ \citenamefont {Calonico}}]{Clivati.2022}%
  \BibitemOpen
  \bibfield  {author} {\bibinfo {author} {\bibfnamefont {C.}~\bibnamefont {Clivati}}, \bibinfo {author} {\bibfnamefont {A.}~\bibnamefont {Meda}}, \bibinfo {author} {\bibfnamefont {S.}~\bibnamefont {Donadello}}, \bibinfo {author} {\bibfnamefont {S.}~\bibnamefont {Virz{\`i}}}, \bibinfo {author} {\bibfnamefont {M.}~\bibnamefont {Genovese}}, \bibinfo {author} {\bibfnamefont {F.}~\bibnamefont {Levi}}, \bibinfo {author} {\bibfnamefont {A.}~\bibnamefont {Mura}}, \bibinfo {author} {\bibfnamefont {M.}~\bibnamefont {Pittaluga}}, \bibinfo {author} {\bibfnamefont {Z.}~\bibnamefont {Yuan}}, \bibinfo {author} {\bibfnamefont {A.~J.}\ \bibnamefont {Shields}}, \bibinfo {author} {\bibfnamefont {M.}~\bibnamefont {Lucamarini}}, \bibinfo {author} {\bibfnamefont {I.~P.}\ \bibnamefont {Degiovanni}},\ and\ \bibinfo {author} {\bibfnamefont {D.}~\bibnamefont {Calonico}},\ }\bibfield  {title} {\bibinfo {title} {Coherent phase transfer for real-world twin-field quantum key distribution},\ }\href@noop {} {\bibfield  {journal} {\bibinfo
  {journal} {Nature Communications}\ }\textbf {\bibinfo {volume} {13}},\ \bibinfo {pages} {157} (\bibinfo {year} {2022})}\BibitemShut {NoStop}%
\bibitem [{\citenamefont {Zhou}\ \emph {et~al.}(2023)\citenamefont {Zhou}, \citenamefont {Lin}, \citenamefont {Jing},\ and\ \citenamefont {Yuan}}]{Zhou.2023}%
  \BibitemOpen
  \bibfield  {author} {\bibinfo {author} {\bibfnamefont {L.}~\bibnamefont {Zhou}}, \bibinfo {author} {\bibfnamefont {J.}~\bibnamefont {Lin}}, \bibinfo {author} {\bibfnamefont {Y.}~\bibnamefont {Jing}},\ and\ \bibinfo {author} {\bibfnamefont {Z.}~\bibnamefont {Yuan}},\ }\bibfield  {title} {\bibinfo {title} {Twin-field quantum key distribution without optical frequency dissemination},\ }\href@noop {} {\bibfield  {journal} {\bibinfo  {journal} {Nature Communications}\ }\textbf {\bibinfo {volume} {14}},\ \bibinfo {pages} {928} (\bibinfo {year} {2023})}\BibitemShut {NoStop}%
\bibitem [{\citenamefont {Xu}\ \emph {et~al.}(2020{\natexlab{b}})\citenamefont {Xu}, \citenamefont {Ma}, \citenamefont {Zhang}, \citenamefont {Lo},\ and\ \citenamefont {Pan}}]{Xu.2020b}%
  \BibitemOpen
  \bibfield  {author} {\bibinfo {author} {\bibfnamefont {F.}~\bibnamefont {Xu}}, \bibinfo {author} {\bibfnamefont {X.}~\bibnamefont {Ma}}, \bibinfo {author} {\bibfnamefont {Q.}~\bibnamefont {Zhang}}, \bibinfo {author} {\bibfnamefont {H.-K.}\ \bibnamefont {Lo}},\ and\ \bibinfo {author} {\bibfnamefont {J.-W.}\ \bibnamefont {Pan}},\ }\bibfield  {title} {\bibinfo {title} {Secure quantum key distribution with realistic devices},\ }\href@noop {} {\bibfield  {journal} {\bibinfo  {journal} {Reviews of Modern Physics}\ }\textbf {\bibinfo {volume} {92}},\ \bibinfo {pages} {131} (\bibinfo {year} {2020}{\natexlab{b}})}\BibitemShut {NoStop}%
\bibitem [{\citenamefont {Zapatero}\ \emph {et~al.}(2025)\citenamefont {Zapatero}, \citenamefont {Navarrete},\ and\ \citenamefont {Curty}}]{zapatero2025implementation}%
  \BibitemOpen
  \bibfield  {author} {\bibinfo {author} {\bibfnamefont {V.}~\bibnamefont {Zapatero}}, \bibinfo {author} {\bibfnamefont {{\'A}.}~\bibnamefont {Navarrete}},\ and\ \bibinfo {author} {\bibfnamefont {M.}~\bibnamefont {Curty}},\ }\bibfield  {title} {\bibinfo {title} {Implementation security in quantum key distribution},\ }\href@noop {} {\bibfield  {journal} {\bibinfo  {journal} {Advanced Quantum Technologies}\ }\textbf {\bibinfo {volume} {8}},\ \bibinfo {pages} {2300380} (\bibinfo {year} {2025})}\BibitemShut {NoStop}%
\bibitem [{\citenamefont {Jain}\ \emph {et~al.}(2016)\citenamefont {Jain}, \citenamefont {Stiller}, \citenamefont {Khan}, \citenamefont {Elser}, \citenamefont {Marquardt},\ and\ \citenamefont {Leuchs}}]{jain2016attacks}%
  \BibitemOpen
  \bibfield  {author} {\bibinfo {author} {\bibfnamefont {N.}~\bibnamefont {Jain}}, \bibinfo {author} {\bibfnamefont {B.}~\bibnamefont {Stiller}}, \bibinfo {author} {\bibfnamefont {I.}~\bibnamefont {Khan}}, \bibinfo {author} {\bibfnamefont {D.}~\bibnamefont {Elser}}, \bibinfo {author} {\bibfnamefont {C.}~\bibnamefont {Marquardt}},\ and\ \bibinfo {author} {\bibfnamefont {G.}~\bibnamefont {Leuchs}},\ }\bibfield  {title} {\bibinfo {title} {Attacks on practical quantum key distribution systems (and how to prevent them)},\ }\href@noop {} {\bibfield  {journal} {\bibinfo  {journal} {Contemporary Physics}\ }\textbf {\bibinfo {volume} {57}},\ \bibinfo {pages} {366} (\bibinfo {year} {2016})}\BibitemShut {NoStop}%
\bibitem [{\citenamefont {Makarov}\ \emph {et~al.}(2024)\citenamefont {Makarov}, \citenamefont {Abrikosov}, \citenamefont {Chaiwongkhot}, \citenamefont {Fedorov}, \citenamefont {Huang}, \citenamefont {Kiktenko}, \citenamefont {Petrov}, \citenamefont {Ponosova}, \citenamefont {Ruzhitskaya}, \citenamefont {Tayduganov}, \citenamefont {Trefilov},\ and\ \citenamefont {Zaitsev}}]{Makarov.2024}%
  \BibitemOpen
  \bibfield  {author} {\bibinfo {author} {\bibfnamefont {V.}~\bibnamefont {Makarov}}, \bibinfo {author} {\bibfnamefont {A.}~\bibnamefont {Abrikosov}}, \bibinfo {author} {\bibfnamefont {P.}~\bibnamefont {Chaiwongkhot}}, \bibinfo {author} {\bibfnamefont {A.~K.}\ \bibnamefont {Fedorov}}, \bibinfo {author} {\bibfnamefont {A.}~\bibnamefont {Huang}}, \bibinfo {author} {\bibfnamefont {E.}~\bibnamefont {Kiktenko}}, \bibinfo {author} {\bibfnamefont {M.}~\bibnamefont {Petrov}}, \bibinfo {author} {\bibfnamefont {A.}~\bibnamefont {Ponosova}}, \bibinfo {author} {\bibfnamefont {D.}~\bibnamefont {Ruzhitskaya}}, \bibinfo {author} {\bibfnamefont {A.}~\bibnamefont {Tayduganov}}, \bibinfo {author} {\bibfnamefont {D.}~\bibnamefont {Trefilov}},\ and\ \bibinfo {author} {\bibfnamefont {K.}~\bibnamefont {Zaitsev}},\ }\bibfield  {title} {\bibinfo {title} {Preparing a commercial quantum key distribution system for certification against implementation loopholes},\ }\href@noop {} {\bibfield  {journal} {\bibinfo  {journal} {Phys. Rev.
  Appl.}\ }\textbf {\bibinfo {volume} {22}},\ \bibinfo {pages} {044076} (\bibinfo {year} {2024})}\BibitemShut {NoStop}%
\bibitem [{\citenamefont {{Federal Office for Information Security (BSI)}}(2023)}]{bsi2023implementation}%
  \BibitemOpen
  \bibfield  {author} {\bibinfo {author} {\bibnamefont {{Federal Office for Information Security (BSI)}}},\ }\href {https://www.bsi.bund.de/SharedDocs/Downloads/EN/BSI/Publications/Studies/QKD-Systems/QKD-Systems.pdf?__blob=publicationFile\&v=3} {\emph {\bibinfo {title} {Implementation Attacks against {QKD} Systems}}},\ \bibinfo {type} {Tech. Rep.}\ (\bibinfo  {institution} {{Federal Office for Information Security (BSI)}},\ \bibinfo {year} {2023})\ \bibinfo {note} {technical report}\BibitemShut {NoStop}%
\bibitem [{\citenamefont {Lo}\ \emph {et~al.}(2012)\citenamefont {Lo}, \citenamefont {Curty},\ and\ \citenamefont {Qi}}]{Lo.2012}%
  \BibitemOpen
  \bibfield  {author} {\bibinfo {author} {\bibfnamefont {H.-K.}\ \bibnamefont {Lo}}, \bibinfo {author} {\bibfnamefont {M.}~\bibnamefont {Curty}},\ and\ \bibinfo {author} {\bibfnamefont {B.}~\bibnamefont {Qi}},\ }\bibfield  {title} {\bibinfo {title} {Measurement-device-independent quantum key distribution},\ }\href@noop {} {\bibfield  {journal} {\bibinfo  {journal} {Physical Review Letters}\ }\textbf {\bibinfo {volume} {108}},\ \bibinfo {pages} {130503} (\bibinfo {year} {2012})}\BibitemShut {NoStop}%
\bibitem [{\citenamefont {Ye}\ \emph {et~al.}(2003)\citenamefont {Ye}, \citenamefont {Peng}, \citenamefont {Jones}, \citenamefont {Holman}, \citenamefont {Hall}, \citenamefont {Jones}, \citenamefont {Diddams}, \citenamefont {Kitching}, \citenamefont {Bize}, \citenamefont {Bergquist}, \citenamefont {Hollberg}, \citenamefont {Robertsson},\ and\ \citenamefont {Ma}}]{Ye.2003}%
  \BibitemOpen
  \bibfield  {author} {\bibinfo {author} {\bibfnamefont {J.}~\bibnamefont {Ye}}, \bibinfo {author} {\bibfnamefont {J.-L.}\ \bibnamefont {Peng}}, \bibinfo {author} {\bibfnamefont {R.~J.}\ \bibnamefont {Jones}}, \bibinfo {author} {\bibfnamefont {K.~W.}\ \bibnamefont {Holman}}, \bibinfo {author} {\bibfnamefont {J.~L.}\ \bibnamefont {Hall}}, \bibinfo {author} {\bibfnamefont {D.~J.}\ \bibnamefont {Jones}}, \bibinfo {author} {\bibfnamefont {S.~A.}\ \bibnamefont {Diddams}}, \bibinfo {author} {\bibfnamefont {J.}~\bibnamefont {Kitching}}, \bibinfo {author} {\bibfnamefont {S.}~\bibnamefont {Bize}}, \bibinfo {author} {\bibfnamefont {J.~C.}\ \bibnamefont {Bergquist}}, \bibinfo {author} {\bibfnamefont {L.~W.}\ \bibnamefont {Hollberg}}, \bibinfo {author} {\bibfnamefont {L.}~\bibnamefont {Robertsson}},\ and\ \bibinfo {author} {\bibfnamefont {L.-S.}\ \bibnamefont {Ma}},\ }\bibfield  {title} {\bibinfo {title} {Delivery of high-stability optical and microwave frequency standards over an optical fiber network},\ }\href@noop {}
  {\bibfield  {journal} {\bibinfo  {journal} {Journal of the Optical Society of America B}\ }\textbf {\bibinfo {volume} {20}},\ \bibinfo {pages} {1459} (\bibinfo {year} {2003})}\BibitemShut {NoStop}%
\bibitem [{\citenamefont {Comandar}\ \emph {et~al.}(2016)\citenamefont {Comandar}, \citenamefont {Lucamarini}, \citenamefont {Fr{\"o}hlich}, \citenamefont {Dynes}, \citenamefont {Yuan},\ and\ \citenamefont {Shields}}]{Comandar.2016c}%
  \BibitemOpen
  \bibfield  {author} {\bibinfo {author} {\bibfnamefont {L.~C.}\ \bibnamefont {Comandar}}, \bibinfo {author} {\bibfnamefont {M.}~\bibnamefont {Lucamarini}}, \bibinfo {author} {\bibfnamefont {B.}~\bibnamefont {Fr{\"o}hlich}}, \bibinfo {author} {\bibfnamefont {J.~F.}\ \bibnamefont {Dynes}}, \bibinfo {author} {\bibfnamefont {Z.}~\bibnamefont {Yuan}},\ and\ \bibinfo {author} {\bibfnamefont {A.~J.}\ \bibnamefont {Shields}},\ }\bibfield  {title} {\bibinfo {title} {Near perfect mode overlap between independently seeded, gain-switched lasers},\ }\href@noop {} {\bibfield  {journal} {\bibinfo  {journal} {Optics express}\ }\textbf {\bibinfo {volume} {24}},\ \bibinfo {pages} {17849} (\bibinfo {year} {2016})}\BibitemShut {NoStop}%
\bibitem [{\citenamefont {Para{\"\i}so}\ \emph {et~al.}(2021)\citenamefont {Para{\"\i}so}, \citenamefont {Woodward}, \citenamefont {Marangon}, \citenamefont {Lovic}, \citenamefont {Yuan},\ and\ \citenamefont {Shields}}]{paraiso2021advanced}%
  \BibitemOpen
  \bibfield  {author} {\bibinfo {author} {\bibfnamefont {T.~K.}\ \bibnamefont {Para{\"\i}so}}, \bibinfo {author} {\bibfnamefont {R.~I.}\ \bibnamefont {Woodward}}, \bibinfo {author} {\bibfnamefont {D.~G.}\ \bibnamefont {Marangon}}, \bibinfo {author} {\bibfnamefont {V.}~\bibnamefont {Lovic}}, \bibinfo {author} {\bibfnamefont {Z.}~\bibnamefont {Yuan}},\ and\ \bibinfo {author} {\bibfnamefont {A.~J.}\ \bibnamefont {Shields}},\ }\bibfield  {title} {\bibinfo {title} {Advanced laser technology for quantum communications (tutorial review)},\ }\href@noop {} {\bibfield  {journal} {\bibinfo  {journal} {Advanced Quantum Technologies}\ }\textbf {\bibinfo {volume} {4}},\ \bibinfo {pages} {2100062} (\bibinfo {year} {2021})}\BibitemShut {NoStop}%
\bibitem [{\citenamefont {Du}\ \emph {et~al.}(2024)\citenamefont {Du}, \citenamefont {Paraiso}, \citenamefont {Pittaluga}, \citenamefont {Lo}, \citenamefont {Dolphin},\ and\ \citenamefont {Shields}}]{du2024twin}%
  \BibitemOpen
  \bibfield  {author} {\bibinfo {author} {\bibfnamefont {H.}~\bibnamefont {Du}}, \bibinfo {author} {\bibfnamefont {T.~K.}\ \bibnamefont {Paraiso}}, \bibinfo {author} {\bibfnamefont {M.}~\bibnamefont {Pittaluga}}, \bibinfo {author} {\bibfnamefont {Y.~S.}\ \bibnamefont {Lo}}, \bibinfo {author} {\bibfnamefont {J.~A.}\ \bibnamefont {Dolphin}},\ and\ \bibinfo {author} {\bibfnamefont {A.~J.}\ \bibnamefont {Shields}},\ }\bibfield  {title} {\bibinfo {title} {Twin-field quantum key distribution with optical injection locking and phase encoding on-chip},\ }\href@noop {} {\bibfield  {journal} {\bibinfo  {journal} {Optica}\ }\textbf {\bibinfo {volume} {11}},\ \bibinfo {pages} {1385} (\bibinfo {year} {2024})}\BibitemShut {NoStop}%
\bibitem [{\citenamefont {Peng}\ \emph {et~al.}(2025)\citenamefont {Peng}, \citenamefont {Chen}, \citenamefont {Xing}, \citenamefont {Wang}, \citenamefont {Wang}, \citenamefont {Liu},\ and\ \citenamefont {Huang}}]{peng2025practical}%
  \BibitemOpen
  \bibfield  {author} {\bibinfo {author} {\bibfnamefont {Q.}~\bibnamefont {Peng}}, \bibinfo {author} {\bibfnamefont {J.-P.}\ \bibnamefont {Chen}}, \bibinfo {author} {\bibfnamefont {T.}~\bibnamefont {Xing}}, \bibinfo {author} {\bibfnamefont {D.}~\bibnamefont {Wang}}, \bibinfo {author} {\bibfnamefont {Y.}~\bibnamefont {Wang}}, \bibinfo {author} {\bibfnamefont {Y.}~\bibnamefont {Liu}},\ and\ \bibinfo {author} {\bibfnamefont {A.}~\bibnamefont {Huang}},\ }\bibfield  {title} {\bibinfo {title} {Practical security of twin-field quantum key distribution with optical phase-locked loop under wavelength-switching attack},\ }\href@noop {} {\bibfield  {journal} {\bibinfo  {journal} {npj Quantum Information}\ }\textbf {\bibinfo {volume} {11}},\ \bibinfo {pages} {7} (\bibinfo {year} {2025})}\BibitemShut {NoStop}%
\bibitem [{\citenamefont {Lucamarini}\ \emph {et~al.}(2015)\citenamefont {Lucamarini}, \citenamefont {Choi}, \citenamefont {Ward}, \citenamefont {Dynes}, \citenamefont {Yuan},\ and\ \citenamefont {Shields}}]{Lucamarini.2015}%
  \BibitemOpen
  \bibfield  {author} {\bibinfo {author} {\bibfnamefont {M.}~\bibnamefont {Lucamarini}}, \bibinfo {author} {\bibfnamefont {I.}~\bibnamefont {Choi}}, \bibinfo {author} {\bibfnamefont {M.~B.}\ \bibnamefont {Ward}}, \bibinfo {author} {\bibfnamefont {J.~F.}\ \bibnamefont {Dynes}}, \bibinfo {author} {\bibfnamefont {Z.~L.}\ \bibnamefont {Yuan}},\ and\ \bibinfo {author} {\bibfnamefont {A.~J.}\ \bibnamefont {Shields}},\ }\bibfield  {title} {\bibinfo {title} {Practical security bounds against the {T}rojan-horse attack in quantum key distribution},\ }\href@noop {} {\bibfield  {journal} {\bibinfo  {journal} {Phys. Rev. X}\ }\textbf {\bibinfo {volume} {5}},\ \bibinfo {pages} {031030} (\bibinfo {year} {2015})}\BibitemShut {NoStop}%
\bibitem [{\citenamefont {Lo}\ \emph {et~al.}(2005)\citenamefont {Lo}, \citenamefont {Ma},\ and\ \citenamefont {Chen}}]{Lo.2005}%
  \BibitemOpen
  \bibfield  {author} {\bibinfo {author} {\bibfnamefont {H.-K.}\ \bibnamefont {Lo}}, \bibinfo {author} {\bibfnamefont {X.}~\bibnamefont {Ma}},\ and\ \bibinfo {author} {\bibfnamefont {K.}~\bibnamefont {Chen}},\ }\bibfield  {title} {\bibinfo {title} {Decoy state quantum key distribution},\ }\href@noop {} {\bibfield  {journal} {\bibinfo  {journal} {Physical Review Letters}\ }\textbf {\bibinfo {volume} {94}},\ \bibinfo {pages} {230504} (\bibinfo {year} {2005})}\BibitemShut {NoStop}%
\bibitem [{\citenamefont {Sajeed}\ \emph {et~al.}(2015)\citenamefont {Sajeed}, \citenamefont {Radchenko}, \citenamefont {Kaiser}, \citenamefont {Bourgoin}, \citenamefont {Pappa}, \citenamefont {Monat}, \citenamefont {Legr\'e},\ and\ \citenamefont {Makarov}}]{sajeed.2015}%
  \BibitemOpen
  \bibfield  {author} {\bibinfo {author} {\bibfnamefont {S.}~\bibnamefont {Sajeed}}, \bibinfo {author} {\bibfnamefont {I.}~\bibnamefont {Radchenko}}, \bibinfo {author} {\bibfnamefont {S.}~\bibnamefont {Kaiser}}, \bibinfo {author} {\bibfnamefont {J.-P.}\ \bibnamefont {Bourgoin}}, \bibinfo {author} {\bibfnamefont {A.}~\bibnamefont {Pappa}}, \bibinfo {author} {\bibfnamefont {L.}~\bibnamefont {Monat}}, \bibinfo {author} {\bibfnamefont {M.}~\bibnamefont {Legr\'e}},\ and\ \bibinfo {author} {\bibfnamefont {V.}~\bibnamefont {Makarov}},\ }\bibfield  {title} {\bibinfo {title} {Attacks exploiting deviation of mean photon number in quantum key distribution and coin tossing},\ }\href@noop {} {\bibfield  {journal} {\bibinfo  {journal} {Phys. Rev. A}\ }\textbf {\bibinfo {volume} {91}},\ \bibinfo {pages} {032326} (\bibinfo {year} {2015})}\BibitemShut {NoStop}%
\bibitem [{\citenamefont {Huang}\ \emph {et~al.}(2019)\citenamefont {Huang}, \citenamefont {Navarrete}, \citenamefont {Sun}, \citenamefont {Chaiwongkhot}, \citenamefont {Curty},\ and\ \citenamefont {Makarov}}]{huang.2019}%
  \BibitemOpen
  \bibfield  {author} {\bibinfo {author} {\bibfnamefont {A.}~\bibnamefont {Huang}}, \bibinfo {author} {\bibfnamefont {A.}~\bibnamefont {Navarrete}}, \bibinfo {author} {\bibfnamefont {S.-H.}\ \bibnamefont {Sun}}, \bibinfo {author} {\bibfnamefont {P.}~\bibnamefont {Chaiwongkhot}}, \bibinfo {author} {\bibfnamefont {M.}~\bibnamefont {Curty}},\ and\ \bibinfo {author} {\bibfnamefont {V.}~\bibnamefont {Makarov}},\ }\bibfield  {title} {\bibinfo {title} {Laser-seeding attack in quantum key distribution},\ }\href@noop {} {\bibfield  {journal} {\bibinfo  {journal} {Phys. Rev. Appl.}\ }\textbf {\bibinfo {volume} {12}},\ \bibinfo {pages} {064043} (\bibinfo {year} {2019})}\BibitemShut {NoStop}%
\bibitem{ThisPaperSM}
  See Supplemental Material for the slow-integrating power meters measurements against the fast intensity modulation (FIM) attack.
\bibitem [{\citenamefont {Tan}\ \emph {et~al.}(2025)\citenamefont {Tan}, \citenamefont {Petrov}, \citenamefont {Zhang}, \citenamefont {Han}, \citenamefont {Liao}, \citenamefont {Makarov}, \citenamefont {Xu},\ and\ \citenamefont {Pan}}]{tan2025wide}%
  \BibitemOpen
  \bibfield  {author} {\bibinfo {author} {\bibfnamefont {H.}~\bibnamefont {Tan}}, \bibinfo {author} {\bibfnamefont {M.}~\bibnamefont {Petrov}}, \bibinfo {author} {\bibfnamefont {W.}~\bibnamefont {Zhang}}, \bibinfo {author} {\bibfnamefont {L.}~\bibnamefont {Han}}, \bibinfo {author} {\bibfnamefont {S.-K.}\ \bibnamefont {Liao}}, \bibinfo {author} {\bibfnamefont {V.}~\bibnamefont {Makarov}}, \bibinfo {author} {\bibfnamefont {F.}~\bibnamefont {Xu}},\ and\ \bibinfo {author} {\bibfnamefont {J.-W.}\ \bibnamefont {Pan}},\ }\bibfield  {title} {\bibinfo {title} {Wide-spectrum security of quantum key distribution},\ }\href@noop {} {\bibfield  {journal} {\bibinfo  {journal} {PRX Quantum}\ }\textbf {\bibinfo {volume} {6}},\ \bibinfo {pages} {040331} (\bibinfo {year} {2025})}\BibitemShut {NoStop}%
\bibitem [{\citenamefont {Rogalski}(2011)}]{rogalski2011infrared}%
  \BibitemOpen
  \bibfield  {author} {\bibinfo {author} {\bibfnamefont {A.}~\bibnamefont {Rogalski}},\ }\href@noop {} {\emph {\bibinfo {title} {Infrared Detectors}}},\ \bibinfo {edition} {2nd}\ ed.\ (\bibinfo  {publisher} {CRC Press},\ \bibinfo {address} {Boca Raton, FL, USA},\ \bibinfo {year} {2011})\BibitemShut {NoStop}%
\bibitem [{\citenamefont {Carr}\ \emph {et~al.}(2003)\citenamefont {Carr}, \citenamefont {Radousky},\ and\ \citenamefont {Demos}}]{Carr.2003}%
  \BibitemOpen
  \bibfield  {author} {\bibinfo {author} {\bibfnamefont {C.~W.}\ \bibnamefont {Carr}}, \bibinfo {author} {\bibfnamefont {H.~B.}\ \bibnamefont {Radousky}},\ and\ \bibinfo {author} {\bibfnamefont {S.~G.}\ \bibnamefont {Demos}},\ }\bibfield  {title} {\bibinfo {title} {Wavelength dependence of laser-induced damage: Determining the damage initiation mechanisms},\ }\href@noop {} {\bibfield  {journal} {\bibinfo  {journal} {Phys. Rev. Lett.}\ }\textbf {\bibinfo {volume} {91}},\ \bibinfo {pages} {127402} (\bibinfo {year} {2003})}\BibitemShut {NoStop}%
\bibitem [{\citenamefont {Palais}(2005)}]{Palais2005}%
  \BibitemOpen
  \bibfield  {author} {\bibinfo {author} {\bibfnamefont {J.~C.}\ \bibnamefont {Palais}},\ }\href@noop {} {\emph {\bibinfo {title} {Fiber Optic Communications}}},\ \bibinfo {edition} {5th}\ ed.\ (\bibinfo  {publisher} {Pearson/Prentice Hall},\ \bibinfo {year} {2005})\BibitemShut {NoStop}%
\bibitem [{\citenamefont {Currás-Lorenzo}\ \emph {et~al.}(2025)\citenamefont {Currás-Lorenzo}, \citenamefont {Navarrete}, \citenamefont {Núñez-Bon}, \citenamefont {Pereira},\ and\ \citenamefont {Curty}}]{Curras-Lorenzo.2025a}%
  \BibitemOpen
  \bibfield  {author} {\bibinfo {author} {\bibfnamefont {G.}~\bibnamefont {Currás-Lorenzo}}, \bibinfo {author} {\bibfnamefont {A.}~\bibnamefont {Navarrete}}, \bibinfo {author} {\bibfnamefont {J.}~\bibnamefont {Núñez-Bon}}, \bibinfo {author} {\bibfnamefont {M.}~\bibnamefont {Pereira}},\ and\ \bibinfo {author} {\bibfnamefont {M.}~\bibnamefont {Curty}},\ }\bibfield  {title} {\bibinfo {title} {Numerical security analysis for quantum key distribution with partial state characterization},\ }\href@noop {} {\bibfield  {journal} {\bibinfo  {journal} {Quantum Science and Technology}\ }\textbf {\bibinfo {volume} {10}},\ \bibinfo {pages} {035031} (\bibinfo {year} {2025})}\BibitemShut {NoStop}%
\bibitem [{\citenamefont {Tamaki}\ \emph {et~al.}(2016)\citenamefont {Tamaki}, \citenamefont {Curty},\ and\ \citenamefont {Lucamarini}}]{Tamaki.2016}%
  \BibitemOpen
  \bibfield  {author} {\bibinfo {author} {\bibfnamefont {K.}~\bibnamefont {Tamaki}}, \bibinfo {author} {\bibfnamefont {M.}~\bibnamefont {Curty}},\ and\ \bibinfo {author} {\bibfnamefont {M.}~\bibnamefont {Lucamarini}},\ }\bibfield  {title} {\bibinfo {title} {Decoy-state quantum key distribution with a leaky source},\ }\href@noop {} {\bibfield  {journal} {\bibinfo  {journal} {New Journal of Physics}\ }\textbf {\bibinfo {volume} {18}},\ \bibinfo {pages} {065008} (\bibinfo {year} {2016})}\BibitemShut {NoStop}%
\bibitem [{\citenamefont {Wang}\ \emph {et~al.}(2018{\natexlab{b}})\citenamefont {Wang}, \citenamefont {Tamaki},\ and\ \citenamefont {Curty}}]{Wang.2018.leaky}%
  \BibitemOpen
  \bibfield  {author} {\bibinfo {author} {\bibfnamefont {W.}~\bibnamefont {Wang}}, \bibinfo {author} {\bibfnamefont {K.}~\bibnamefont {Tamaki}},\ and\ \bibinfo {author} {\bibfnamefont {M.}~\bibnamefont {Curty}},\ }\bibfield  {title} {\bibinfo {title} {Finite-key security analysis for quantum key distribution with leaky sources},\ }\href@noop {} {\bibfield  {journal} {\bibinfo  {journal} {New Journal of Physics}\ }\textbf {\bibinfo {volume} {20}},\ \bibinfo {pages} {083027} (\bibinfo {year} {2018}{\natexlab{b}})}\BibitemShut {NoStop}%
\bibitem [{\citenamefont {Navarrete}\ and\ \citenamefont {Curty}(2022)}]{Navarrete.2022}%
  \BibitemOpen
  \bibfield  {author} {\bibinfo {author} {\bibfnamefont {A.}~\bibnamefont {Navarrete}}\ and\ \bibinfo {author} {\bibfnamefont {M.}~\bibnamefont {Curty}},\ }\bibfield  {title} {\bibinfo {title} {Improved finite-key security analysis of quantum key distribution against trojan-horse attacks},\ }\href@noop {} {\bibfield  {journal} {\bibinfo  {journal} {Quantum Science and Technology}\ }\textbf {\bibinfo {volume} {7}},\ \bibinfo {pages} {035021} (\bibinfo {year} {2022})}\BibitemShut {NoStop}%
\bibitem [{\citenamefont {Curr{\'a}s-Lorenzo}\ \emph {et~al.}(2025)\citenamefont {Curr{\'a}s-Lorenzo}, \citenamefont {Pereira}, \citenamefont {Kato}, \citenamefont {Curty},\ and\ \citenamefont {Tamaki}}]{curras-lorenzo.2025b}%
  \BibitemOpen
  \bibfield  {author} {\bibinfo {author} {\bibfnamefont {G.}~\bibnamefont {Curr{\'a}s-Lorenzo}}, \bibinfo {author} {\bibfnamefont {M.}~\bibnamefont {Pereira}}, \bibinfo {author} {\bibfnamefont {G.}~\bibnamefont {Kato}}, \bibinfo {author} {\bibfnamefont {M.}~\bibnamefont {Curty}},\ and\ \bibinfo {author} {\bibfnamefont {K.}~\bibnamefont {Tamaki}},\ }\bibfield  {title} {\bibinfo {title} {Security framework for quantum key distribution with imperfect sources},\ }\href@noop {} {\bibfield  {journal} {\bibinfo  {journal} {Optica Quantum}\ }\textbf {\bibinfo {volume} {3}},\ \bibinfo {pages} {525} (\bibinfo {year} {2025})}\BibitemShut {NoStop}%
\bibitem [{\citenamefont {Sixto}\ \emph {et~al.}(2025)\citenamefont {Sixto}, \citenamefont {Navarrete}, \citenamefont {Pereira}, \citenamefont {Curr{\'a}s-Lorenzo}, \citenamefont {Tamaki},\ and\ \citenamefont {Curty}}]{sixto2025quantum}%
  \BibitemOpen
  \bibfield  {author} {\bibinfo {author} {\bibfnamefont {X.}~\bibnamefont {Sixto}}, \bibinfo {author} {\bibfnamefont {{\'A}.}~\bibnamefont {Navarrete}}, \bibinfo {author} {\bibfnamefont {M.}~\bibnamefont {Pereira}}, \bibinfo {author} {\bibfnamefont {G.}~\bibnamefont {Curr{\'a}s-Lorenzo}}, \bibinfo {author} {\bibfnamefont {K.}~\bibnamefont {Tamaki}},\ and\ \bibinfo {author} {\bibfnamefont {M.}~\bibnamefont {Curty}},\ }\bibfield  {title} {\bibinfo {title} {Quantum key distribution with imperfectly isolated devices},\ }\href@noop {} {\bibfield  {journal} {\bibinfo  {journal} {Quantum Science and Technology}\ }\textbf {\bibinfo {volume} {10}},\ \bibinfo {pages} {035034} (\bibinfo {year} {2025})}\BibitemShut {NoStop}%
\end{thebibliography}

%

\makeatletter
\renewcommand \thesection{S\@arabic\c@section}
\renewcommand \thetable{S\@arabic\c@table}
\renewcommand \thefigure{S\@arabic\c@figure}
\makeatother
\renewcommand{\theequation}{S.\arabic{equation}}

\clearpage
\onecolumngrid

\setcounter{figure}{0}
\setcounter{equation}{0}
\setcounter{table}{0}
\setcounter{section}{0}
\setcounter{page}{1}





\begin{center}
{\LARGE\textbf{Supplemental Material}}\\[0.3em]
{\LARGE Reference Beam Attacks against Twin-Field Quantum Key Distribution}\\[0.2em]
{\LARGE using Optical Injection Locking}
\end{center}

\vspace{0.5em}

\begin{center}
Sergio Ju\'arez$^*$, Alessandro Marcomini$^*$, Mikhail Petrov, Robert I. Woodward,\\
Toby J. Dowling, R. Mark Stevenson, Marcos Curty, and Davide Rusca

\vspace{0.3em}
{\small $^*$These authors contributed equally. Correspondence: sergio.juarez@toshiba.eu, amarcomini@vqcc.uvigo.es}
\end{center}

\vspace{1.5em}

\setcounter{figure}{0}
\setcounter{equation}{0}
\setcounter{table}{0}
\setcounter{section}{0}

\renewcommand{\thefigure}{S\arabic{figure}}

\section*{Power Meters Measurements}

\begin{figure}[htb]
    \centering
    \includegraphics[width=0.8\linewidth]{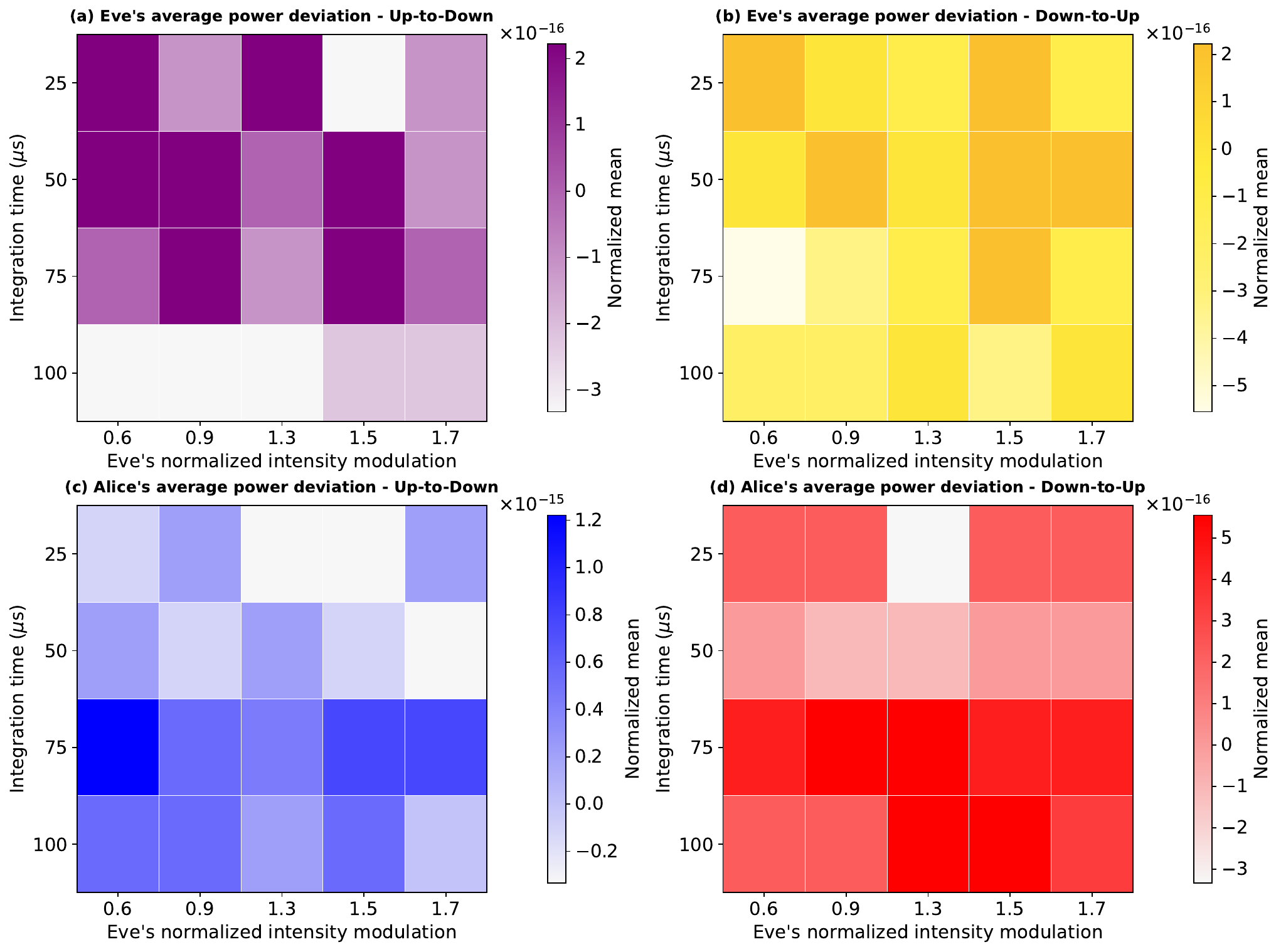}
    \caption{\textbf{Power meters results - deviation of the average optical power under attack, with respect to the attack-free scenario}. These plots display the average recorded optical power of the signal for different integration periods of slow-integrating power meters (without spectral filtering), upon changing the magnitude of the attack.
    The $x$-coordinate corresponds to the normalized peak-to-peak amplitude of the injected modulation (see e.g., Fig.~3(a) of the main text). 
    For each configuration, the average power is normalized to its mean value and compared with the average measurement of the power meter when the modulation is turned off, for each integration time. Results show that, over a few seconds of data acquisition, slow-integrating power meters without spectral filtering are incapable of detecting power oscillations happening on a timescale up to two orders of magnitude smaller than their integration windows.}
    \label{fig:power_meters_avg}
\end{figure}

\begin{figure}[htb]
    \centering
    \includegraphics[width=0.8\linewidth]{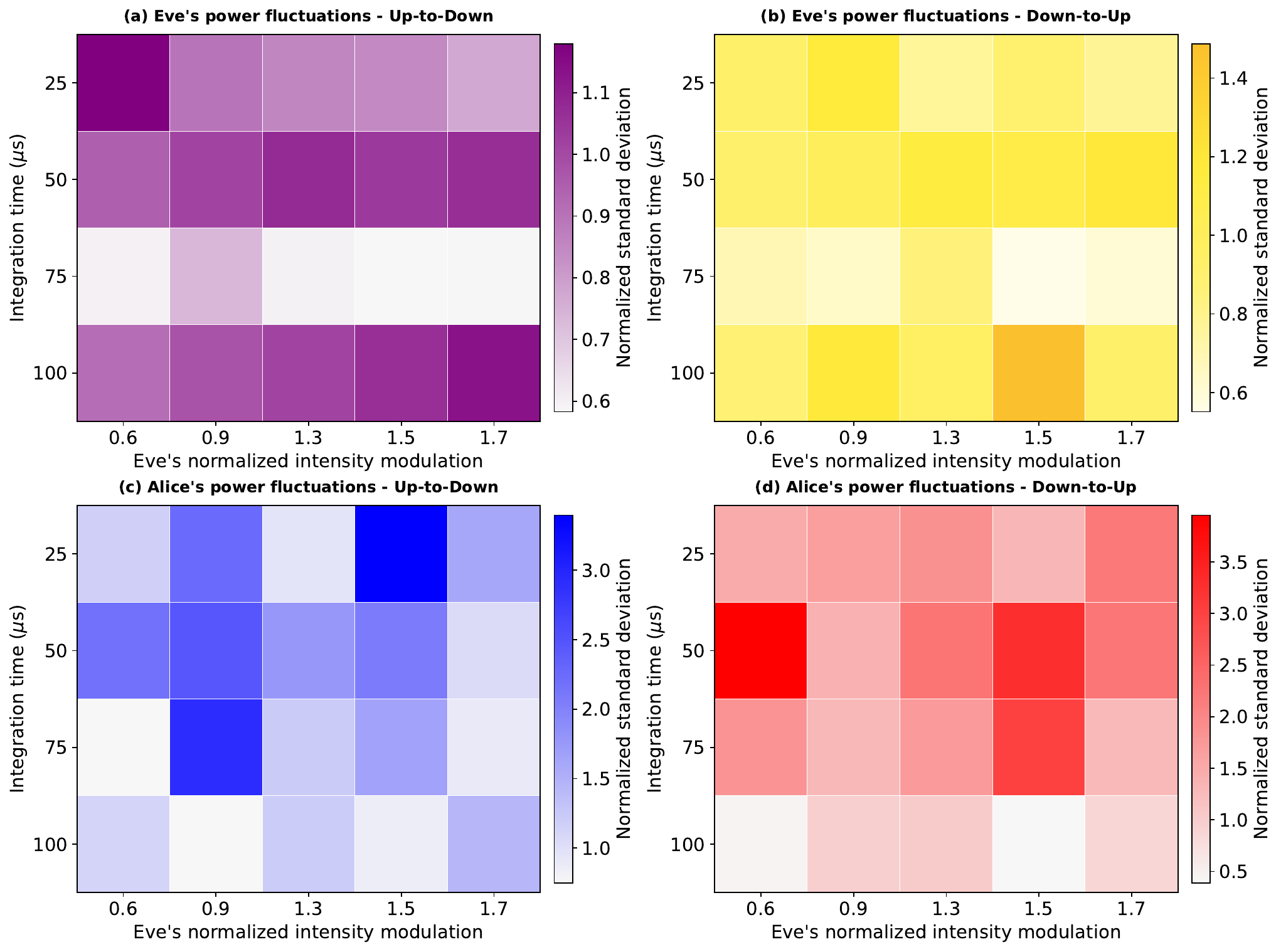}
    \caption{\textbf{Power meters results - power fluctuations under attack, with respect to the attack-free scenario}. These plots display the standard deviation of the recorded optical power for different integration periods of slow-integrating power meters (without spectral filtering), upon changing the magnitude of the attack.
    The $x$-coordinate corresponds to the normalized peak-to-peak amplitude of the injected modulation (see e.g., Fig.~3(a) of the main text). 
    For each configuration, we compute the standard deviation of the optical power normalized to its mean value. Data are reported in units of the standard deviation of the attack-free case, corresponding to the power meter measurements when the modulation is turned off.
    While for some particular parameter combinations the power fluctuations induced by the modulation on Alice's laser surpass the expected one, in most cases power fluctuations are comparable with the attack-free scenario, meaning that the attack can go unnoticed by slow-integrating power meters operating without spectral filtering.}
    \label{fig:power_meters_std}
\end{figure}

Here we report the power meters measurement data, as discussed in the ``Methods" section of the main text. These measurements were performed using slow-integrating power meters (integration times from 25~$\mu$s to 100~ms) without spectral filtering, demonstrating their vulnerability to fast intensity modulation attacks operating at 1~GHz. Figure~\ref{fig:power_meters_avg} shows the deviation of average optical power under attack, while Figure~\ref{fig:power_meters_std} shows the power fluctuations (standard deviation), both normalized to the non-attacked scenario. The results demonstrate that slow-integrating power meters without spectral filtering are insufficient for detecting FIM attacks.

\end{document}